\documentclass[12pt,a4paper]{article}
\usepackage{amssymb,amsmath,graphicx,subcaption,hyperref,cite}
\usepackage{epstopdf}
\usepackage[utf8]{inputenc}
\usepackage{color}
\usepackage[english]{babel}
\begin{document}

\begin{center}
 {\Large\bf Modelling the Spread of an Epidemic in Presence of Vaccination using Cellular Automata}
\end{center}
\vskip 1 cm
\begin{center} %
 Agniva Datta$^1$ and Muktish Acharyya$^2$
  
 \textit{Department of Physics, Presidency University,} 
  
 \textit{86/1 College Street, Kolkata-700073, India} 
 \vskip 0.2 cm
 \textit{Email$^1$:agnivadatta98@gmail.com}
  
 \textit{Email$^2$:muktish.physics@presiuniv.ac.in}
\end{center}
\vspace {1.0 cm}
\vskip 0.5 cm 

\noindent {\large\bf Abstract:}
The results of Kermack-McKendrick SIR model are planned to be reproduced by cellular automata (CA) lattice model. The CA algorithms are proposed to study 
the model of an epidemic, systematically. The basic goal is to capture the effects of spreading of
infection over a scale of length. This CA model can provide the rate of growth
of the infection over the space which was lacking in the mean-field like SIR model. The motion of the circular front of an infected cluster shows a linear behaviour in time. The correlation of a particular site to be infected, with respect to the central site is also studied. 
The outcomes of the CA model are in good agreement with those obtained from SIR model.  The results of vaccination have been also incorporated in the CA algorithm with a satisfactory
degree of success. The advantage of the present model is that it can shed a considerable amount of light on the physical properties of the spread of a typical epidemic in a simple, yet robust way.


\newpage

\section{Introduction}

Computational modelling has become an important area of research in the present
prevailing pandemic situation. Particularly, the spreading of an epidemic, is the main
focus of such studies. The original model of the spread of an epidemic has been proposed in 1927 by
Kermack and McKendrick\cite{strogatz,kermack}. The recent studies in the context of COVID-19, the novel coronavirus (COVID-19), which causes acute respiratory trouble in the human respiratory tract (which may be fatal) has spread
to many countries all over the world and has already been declared as a pandemic by the World Health Organisation\cite{covid1,covid2}. Systematic analyses on the available data of the number of cases and deaths have been attempted recently, and a few data-driven models have also been proposed\cite{gaeta,chen,kastner}. The timing of social policies restricting the social movements is suggested in a recent study\cite{liu1}. The transmission of the disease and spreading by social mixing are studied by computer  simulation\cite{fang}. A novel fractional time delay dynamic system(FTDD) was proposed to describe  the local outbreak of COVID-19\cite{cheng}. The evaluation
of the outbreak in Wuhan, China has been analysed\cite{zhau}. A recent study aimed to establish an early screening model to distinguish COVID-19 pneumonia from Influenza-A viral pneumonia and healthy cases with pulmonary CT images using deep learning techniques\cite{xu}. The Spatio-temporal growth of the COVID-19 outbreak was recently studied\cite{biswas}. The reproduction of data and modelling of the spreading of COVID-19 has
been studied using SIR model on Euclidean network\cite{khaleque}. An interesting phase
transition was reported recently in the Kermack-McKendrick SIR model\cite{datta}.

The studies mentioned above, are mostly on the models based on differential equations suitable to analyse the temporal evolution of the susceptible, infected and removed population. However, these kinds of models cannot give rise to the spatial growth and correlations among different types of populations. 

To study the spatial growth (and spreading of disease) one needs the lattice cellular automata CA \cite{fu} models. The spatial organization and evolution period of epidemic was studied in CA model\cite{liu2}. The pandemic has been compared with the problem of percolation recently\cite{ziff}. The impact of the delay in time has been studied in the CA model of epidemic\cite{sharma}. The critical behaviours have been studied in the SIR lattice model\cite{tome}. The effects of movements of the population and vaccination have been studied in the
CA model\cite{sirakoulis}. Recently, some studies are done on the COVID-19 related pandemic with 
CA modelling. To understand different temporal patterns of infections the CA model has been used\cite{bagchi}. The sequential genetic algorithm based probabilistic Cellular Automata has been used to study the dynamics of COVID-19\cite{saumik}. The impact of social isolation in the spreading of COVID-19 in Brazil, was also studied by using CA model\cite{schimit}. In the context of reduction of the probability of spreading of infections, it may be mentioned that the limited number
of immunization units can considerably suppress the probability of infection\cite{hans}.

In this article, we propose a realistic cellular automata model (based on discrete automaton and specified rule) of epidemic, which incorporates persistence (after getting infected) of the infected population.
Moreover, the removed population (in the SIR model) has been efficiently described here by distinguishing between recovered and dead populations. The effect of immunity through vaccination is also a major development of this model. The paper is organised as follows. The lattice model (both for without vaccination and with vaccination) is described in the next section (section-2). Section-3 is devoted to the numerical/simulational results (here also, both for without vaccination and with vaccination).
The paper ends with a few concluding remarks in Section-4.

\section{Lattice Model of Epidemic}
\begin{figure}[h]
 \centering
 \includegraphics[width=0.88\columnwidth]{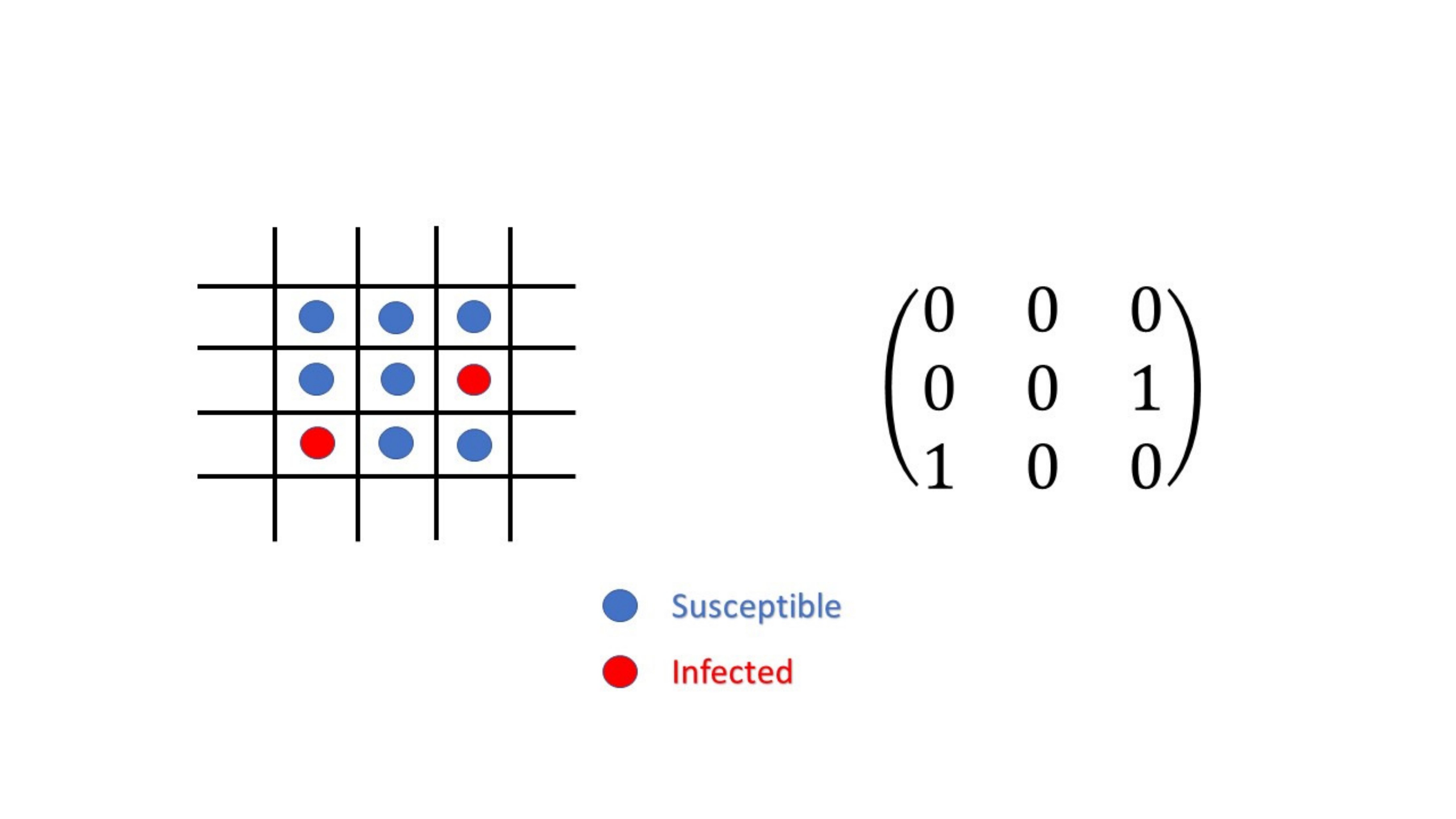}
 \caption{(Color online) The figure shows the lattice along with its matrix counterpart at $t=0$. Blue dots represent the susceptible population and red dots represent the infected population as labelled in the figure.}
 \label{fig:lattice}
\end{figure}

In this model, we consider a two-dimensional lattice arranged in an $N \times N$ grid of cells where each cell corresponds to an individual, such that, there are $N^{2}$ individuals out of which $K$ random individuals are infected and the rest are susceptible to infection. We treat this mathematically by considering an $N \times N$ null matrix $A$ whose all elements are 0 (corresponding to susceptible population) and then randomly set $K$ values to 1 (corresponding to infected population), as shown in FIG:~\ref{fig:lattice}. This corresponds to the lattice at the initial generation of time ($t=0$). Now we evolve the lattice and therefore the matrix using two sets of rules, the first one in absence of vaccination and the latter one, in presence of vaccination. 

In this model, we have considered Von-Neumann neighbourhood (only four nearest neighbours of a particular central cell). One can also consider, Moore neighbourhood
(eight neighbours of a central cell) in this kind of study. Considering more neighbours
will increase the rate of infection. Here, we just tried to model the facts with minimal
interaction.

Here, the boundary condition is considered as open, since the model aims to simulate the spread of the epidemic in a region where the border is considered almost sealed (negligible chance of acquiring infection from outside). Here, the periodic boundary condition is meaningless.

\subsection{In Absence of Vaccination}
First, we label the susceptible population with the number $0$, the infected population with a range of values from $1$ to $2$, corresponding to the infection period of the virus in the host, the recovered population with the number $2$ and the dead population with the number $-1$. Now, the rules are: \par
\begin{figure}[h]
 \centering
 \includegraphics[width=0.88\columnwidth]{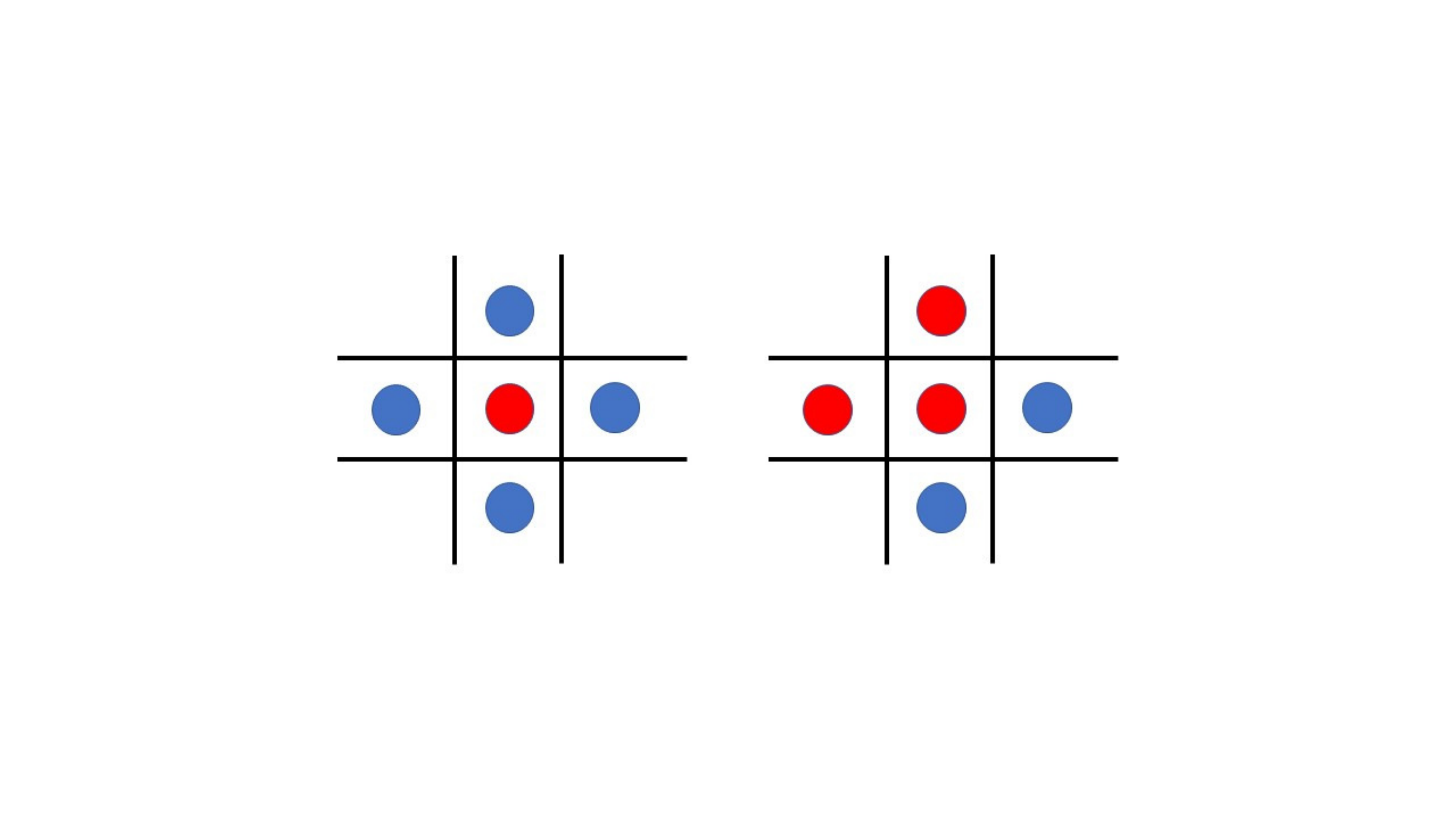}
 \caption{(Color online) Each of the four blue (susceptible) neighbours have p0 probability of acquiring infection from the central red (infected) cell as shown in the picture on left. If one or more of the neighbours are red (infected), they cannot get infected from the central cell, i.e., only the two blue neighbours in the picture on right have chance of getting infection independently with a probability p0.}
 \label{fig:infection}
\end{figure}

 \vskip 0.5cm
(1) If a particular central cell is infected, each of its susceptible nearest neighbours have a probability p0 of getting infected independently. If one or more neighbouring cells are infected, they remain unaffected by the central cell, as shown in FIG:~\ref{fig:infection}. We treat this computationally using the following algorithm:  \vskip 0.5cm \par

For any $t=t'>0$ and for all $i,j < N$, if $1 \leq A[i,j] < 2 $, 
\par \vskip 0.35cm \hskip 0.5 cm
(i) If $A[i-1,j]$ = 0, set $r$ = random (generate a random number between 0 and 1). 
\par \vskip 0.25cm \hskip 1.1cm 
If $r< p0$, set $A[i-1,j]$ = 1, \hskip0.2cm  at $t=t'+1$. 
\par  \vskip 0.35cm \hskip 0.5cm
(ii) Repeat step-(i) for $A[i+1,j]$, $A[i,j-1]$ and $A[i,j+1]$ respectively in place of $A[i-1,j]$.
\par
\begin{figure}[h]
 \centering
 \includegraphics[width=0.88\columnwidth]{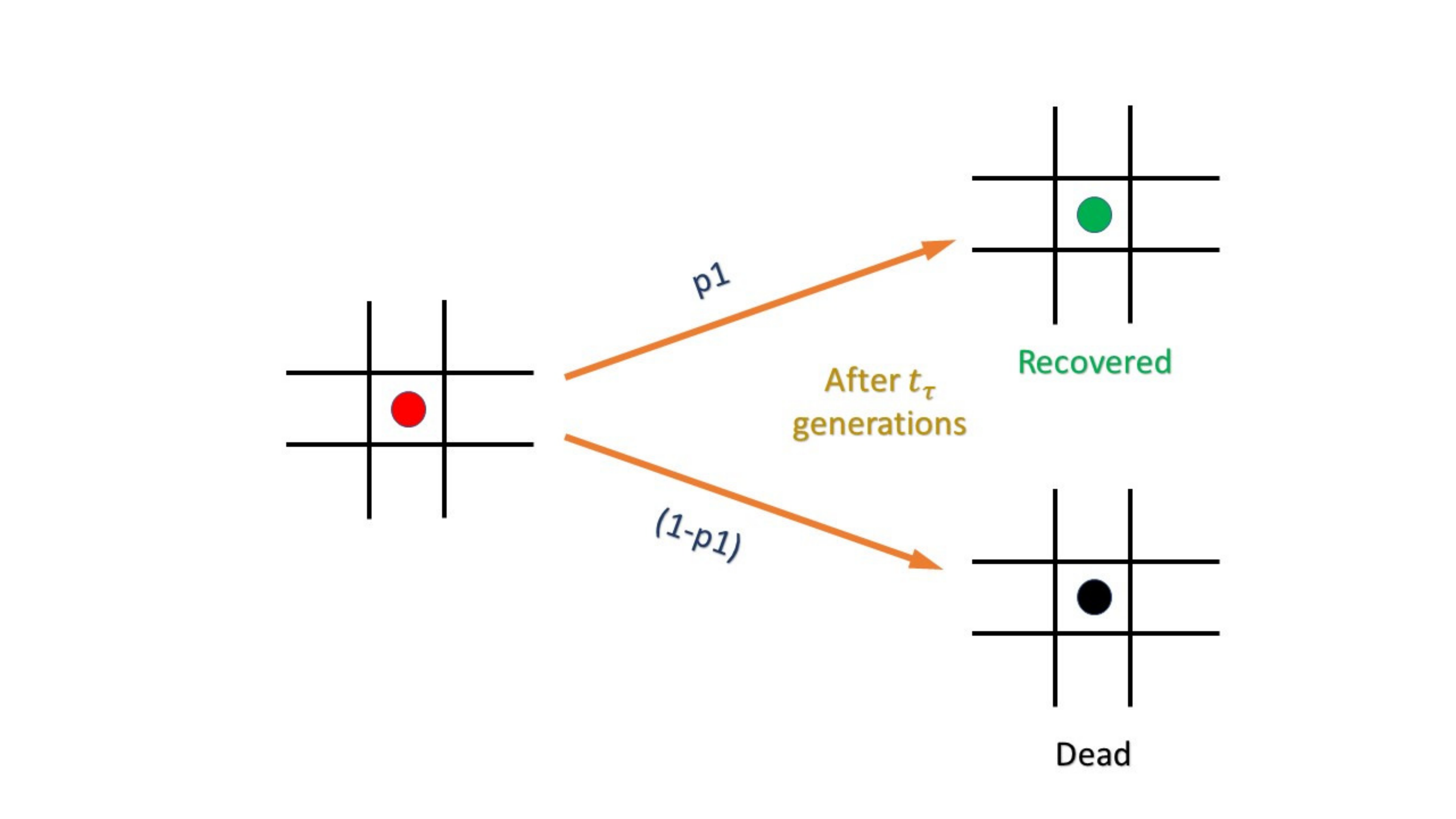}
 \caption{(Color online) When a particular cell gets infected, it stays infected for $t_{\tau}$ generations, which actually serves as the infection period of the virus. After that it either gets recovered with a probability p1 or dies with a probability (1-p1).}
 \label{fig:removal}
\end{figure}

 \vskip 0.5cm
(2) If a particular cell gets infected, it remains infected for the next $t_{\tau}$ generations. After that, it turns into removed population as shown in FIG:~\ref{fig:removal}, i.e., \vskip 0.35cm \par   a] It either gets recovered with a probability p1. In that case, it remains mildly susceptible. \par b] It dies with a probability (1-p1) and remains as it is in the upcoming generations. \par  Computationally, we can implement this as follows: 

\par 
(i) For any $t=t'$ and for all $i,j < N$, if $1 \leq A[i,j] < 2-\frac{1}{t_{\tau}} $, set \par \vskip 0.25cm  \hskip 1.1cm
        {$A[i,j] = A[i,j] + \frac{1}{t_{\tau}}$, \hskip 0.2cm at $t=t'+1$.}

\par \vskip 0.35cm

(ii) For any $t=t'>0$ and for all $i,j < N$, if $A[i,j] = 2-\frac{1}{t_{\tau}}$, set 
\par \vskip 0.25cm \hskip 1.1cm
$r$ = random (generate a random number between 0 an 1).
\par \vskip 0.25cm \hskip 1.1cm 
If $r < p1$, set $A[i,j]$ = 2, \hskip0.2cm at $t=t'+1$. 
\par \vskip 0.25cm \hskip 1.1cm
Else, set $A[i,j]$ = -1, \hskip0.2cm at $t=t'+1$. Here, we set $t_{\tau}=4$. \vskip 0.5cm

(3) If a particular cell is recovered and it lies in the neighbourhood of an infected cell, it has a p2 probability of getting re-infected. The mode of infection is same as that, discussed in Rule (1). We implement this mathematically in the following way: 
\vskip 0.35cm \par

For any $t=t'>0$ and for all $i,j < N$, if $1 \leq A[i,j] < 2 $, 
\par \vskip 0.35cm \hskip 0.5 cm
(i) If $A[i-1,j]$ = 2, set $r$ = random (generate a random number between 0 and 1). 
\par \vskip 0.25cm \hskip 1.1cm 
If $r< p2$, set $A[i-1,j]$ = 1, \hskip0.2cm  at $t=t'+1$. 
\par  \vskip 0.35cm \hskip 0.5cm
(ii) Repeat step-(i) for $A[i+1,j]$, $A[i,j-1]$ and $A[i,j+1]$ respectively in place of $A[i-1,j]$.

\par \vskip 0.5cm

We run the cellular automata simulation following the rules as discussed above for a 100x100 lattice and a 500x500 lattice using the parameters as listed below in Table:~\ref{table1}. All the parameters are chosen in accordance with the COVID-19 epidemic which has moderate infection rate and a low fatality rate\cite{rate,fatality}. \par
\begin{figure*}[htpb]
 \centering
 (a)\includegraphics[width=0.46\columnwidth]{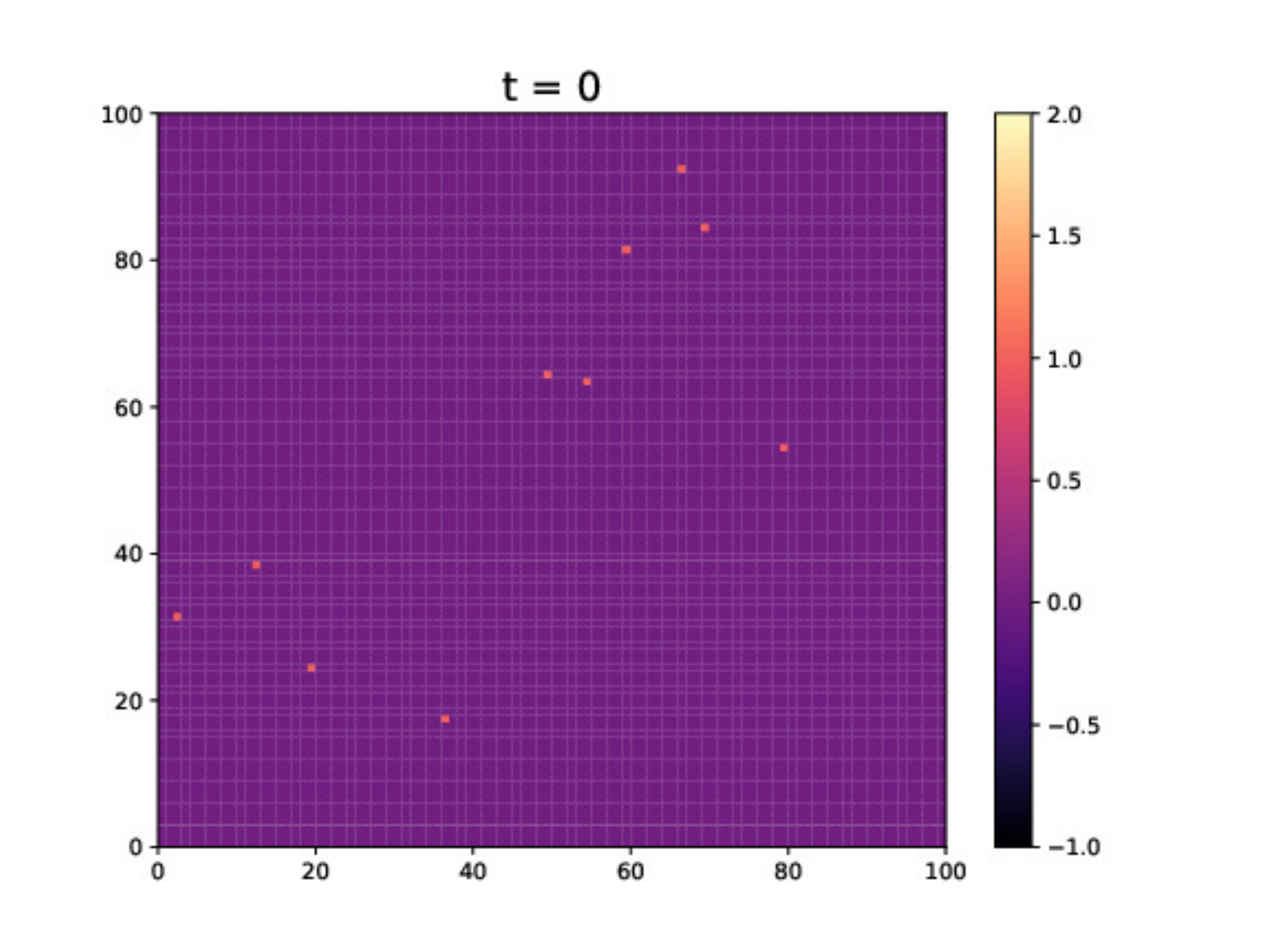}
 (b)\includegraphics[width=0.46\columnwidth]{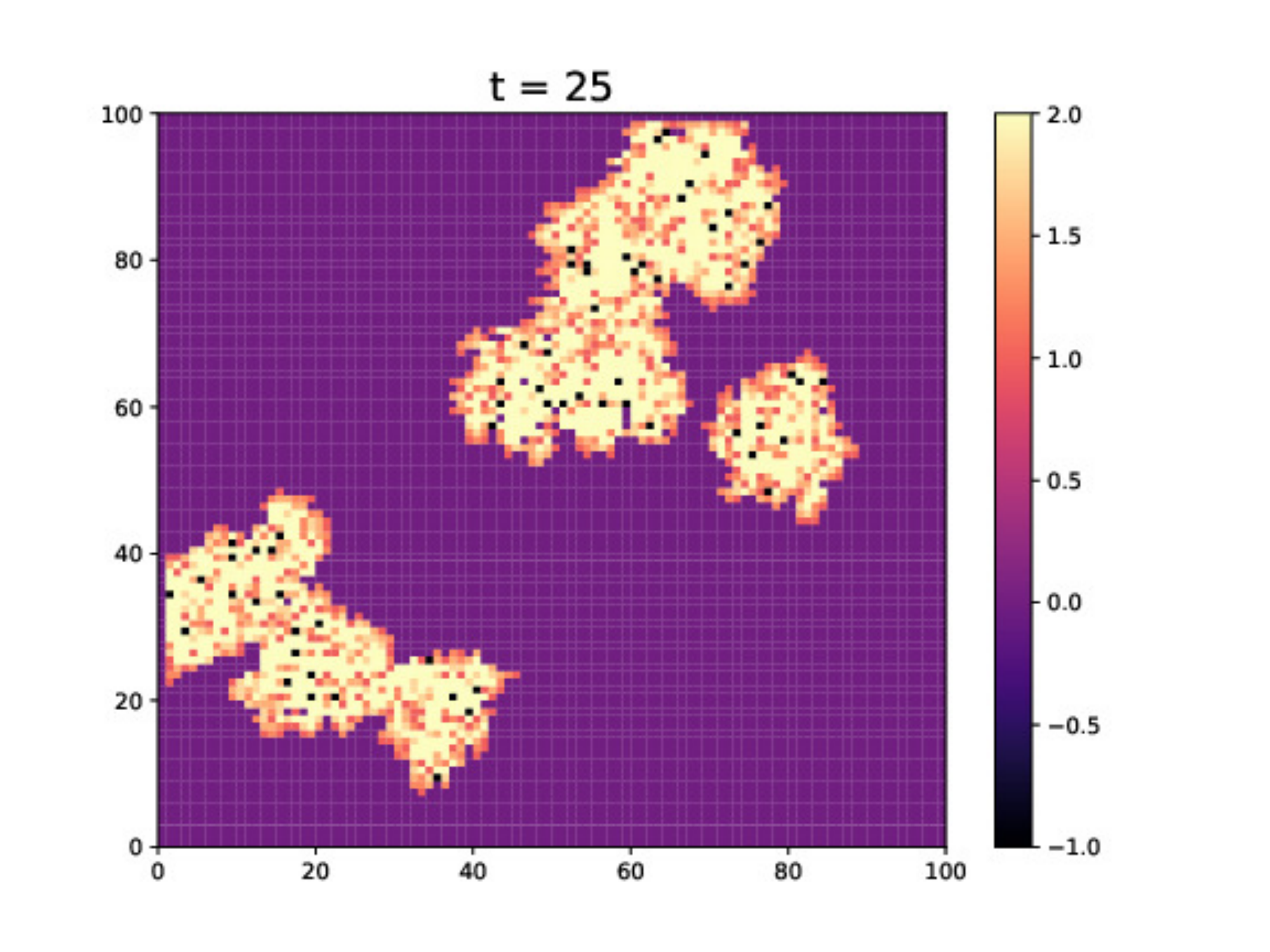}
 \\
 (c)\includegraphics[width=0.46\columnwidth]{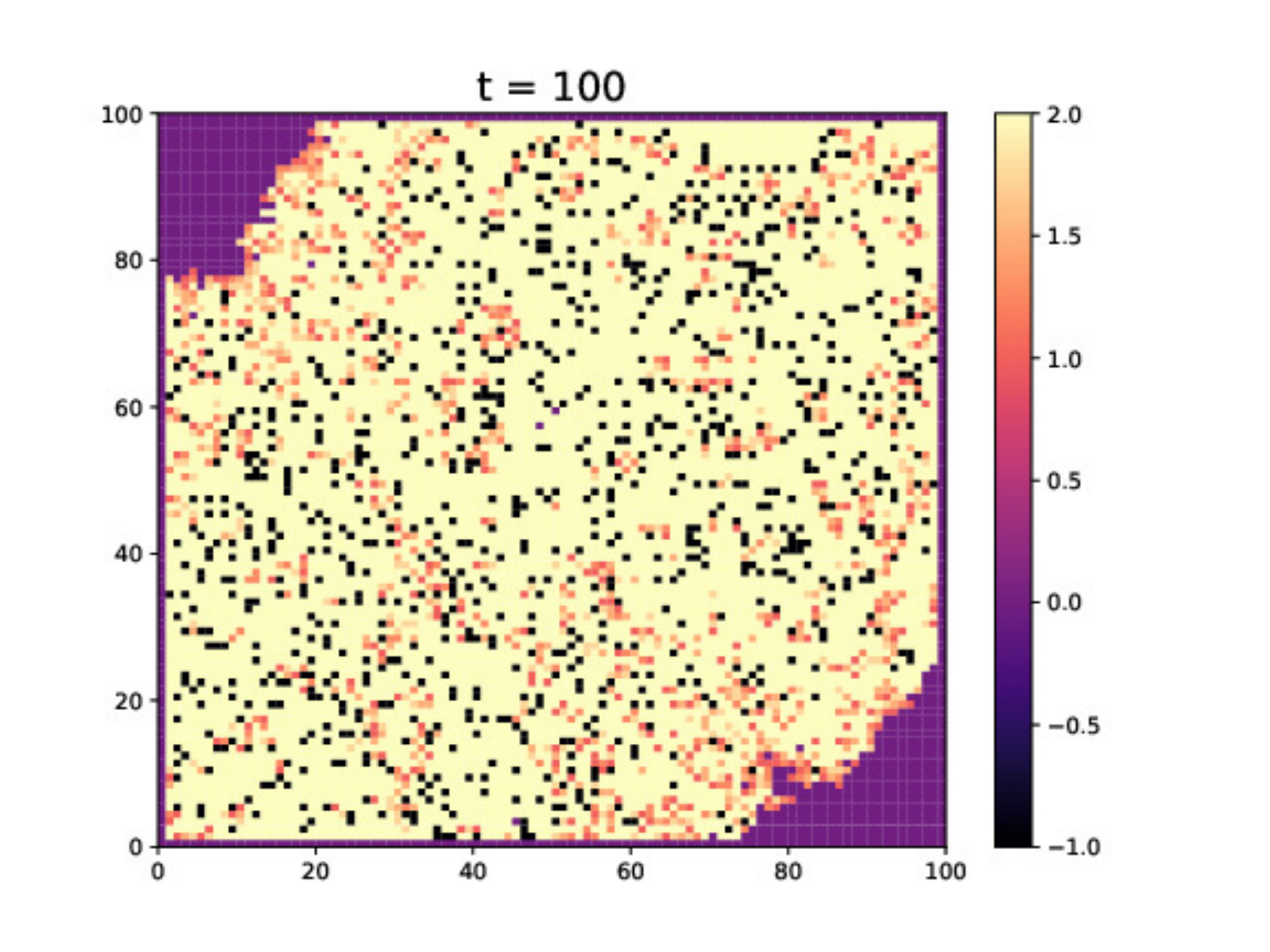}
 (d)\includegraphics[width=0.46\columnwidth]{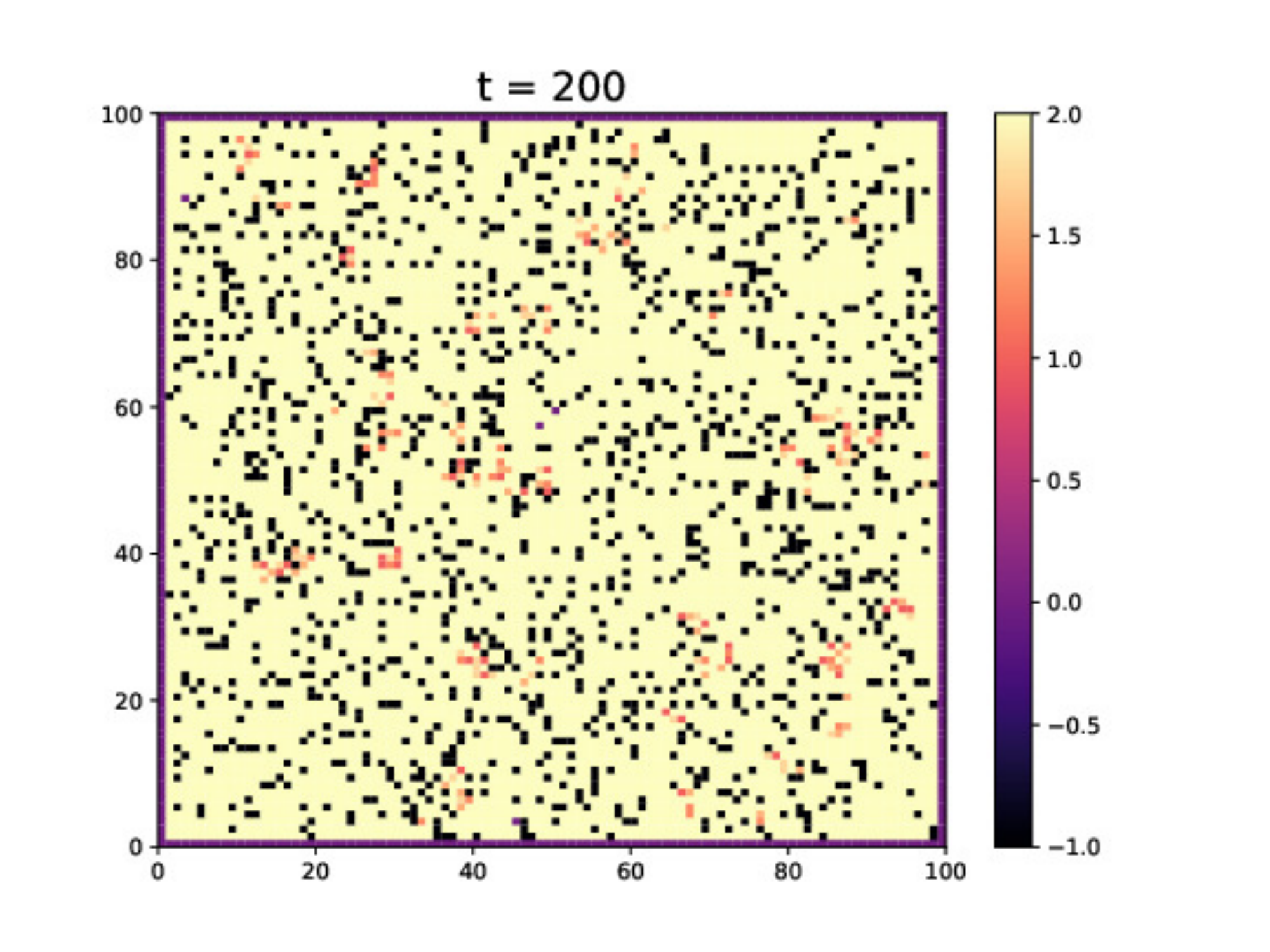}
 \caption{ (Color online) Chosen Parameters: Total population = 10000; total number of initially infected population = 10. The number -1 (black) represents the dead population; the number 0 (violet) represents the susceptible population; the range 1-2 (shades of red) represents the infected population; the number 2 (cream) represents the recovered population. Four frames of the Lattice Model Simulation in absence of vaccination are shown at    (a) $t=0$ (b) $t=25$ (c) $t=100$ (d) $t=200$. }
 \label{fig:100novac}
\end{figure*}

\begin{figure*}[htpb]
 \centering
 (a)\includegraphics[width=0.46\columnwidth]{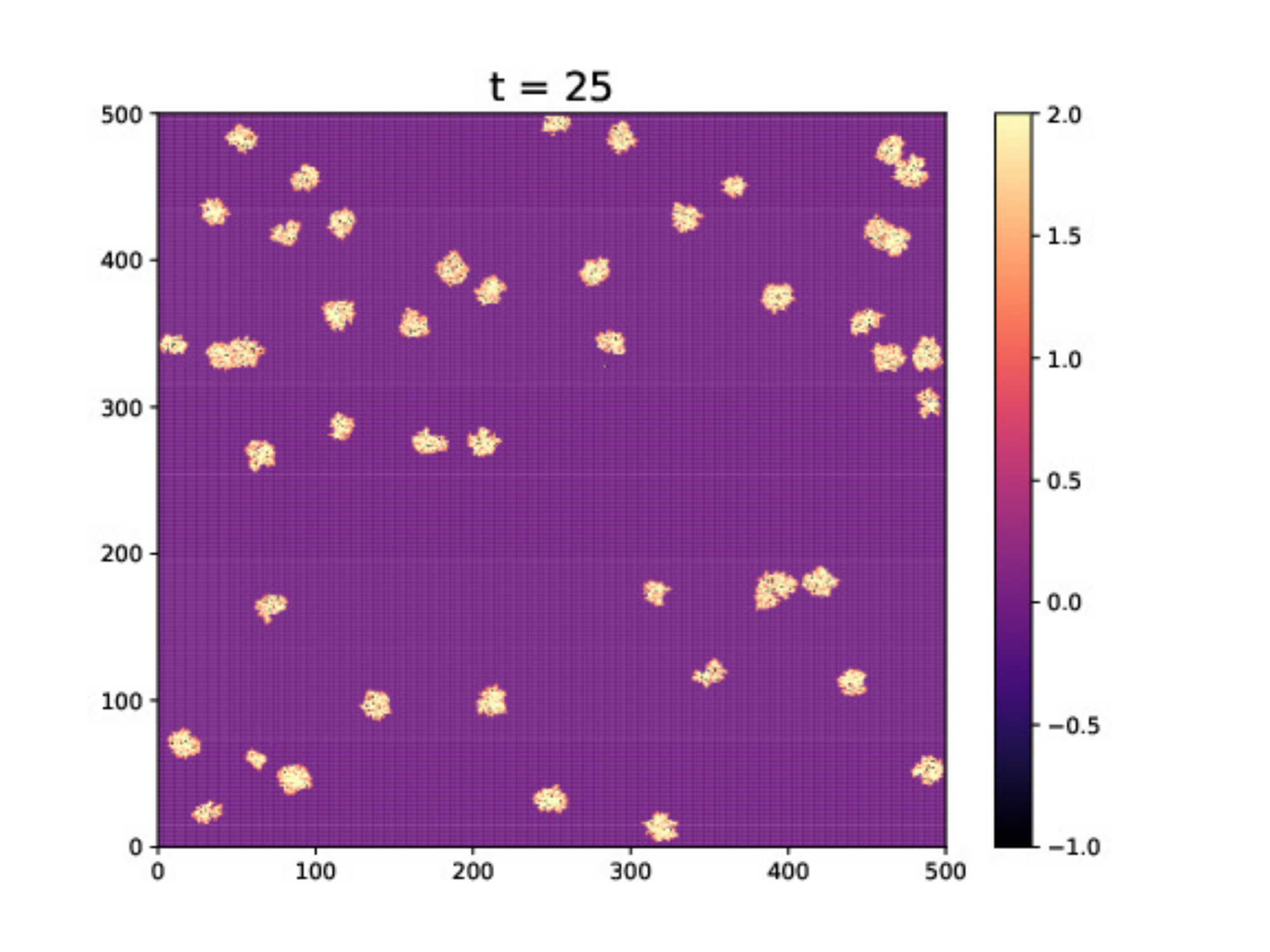}
 (b)\includegraphics[width=0.46\columnwidth]{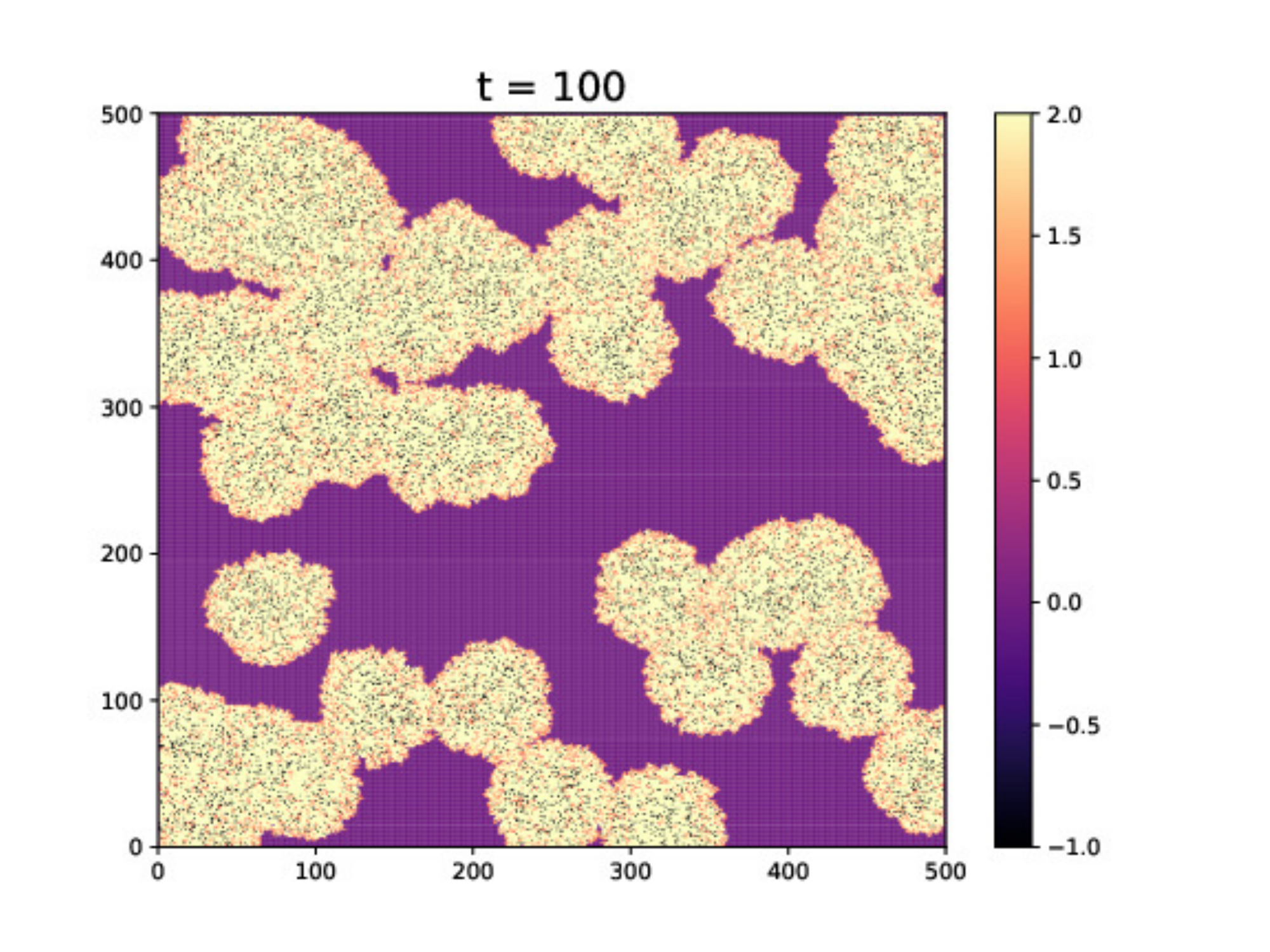}
 \\
 (c)\includegraphics[width=0.46\columnwidth]{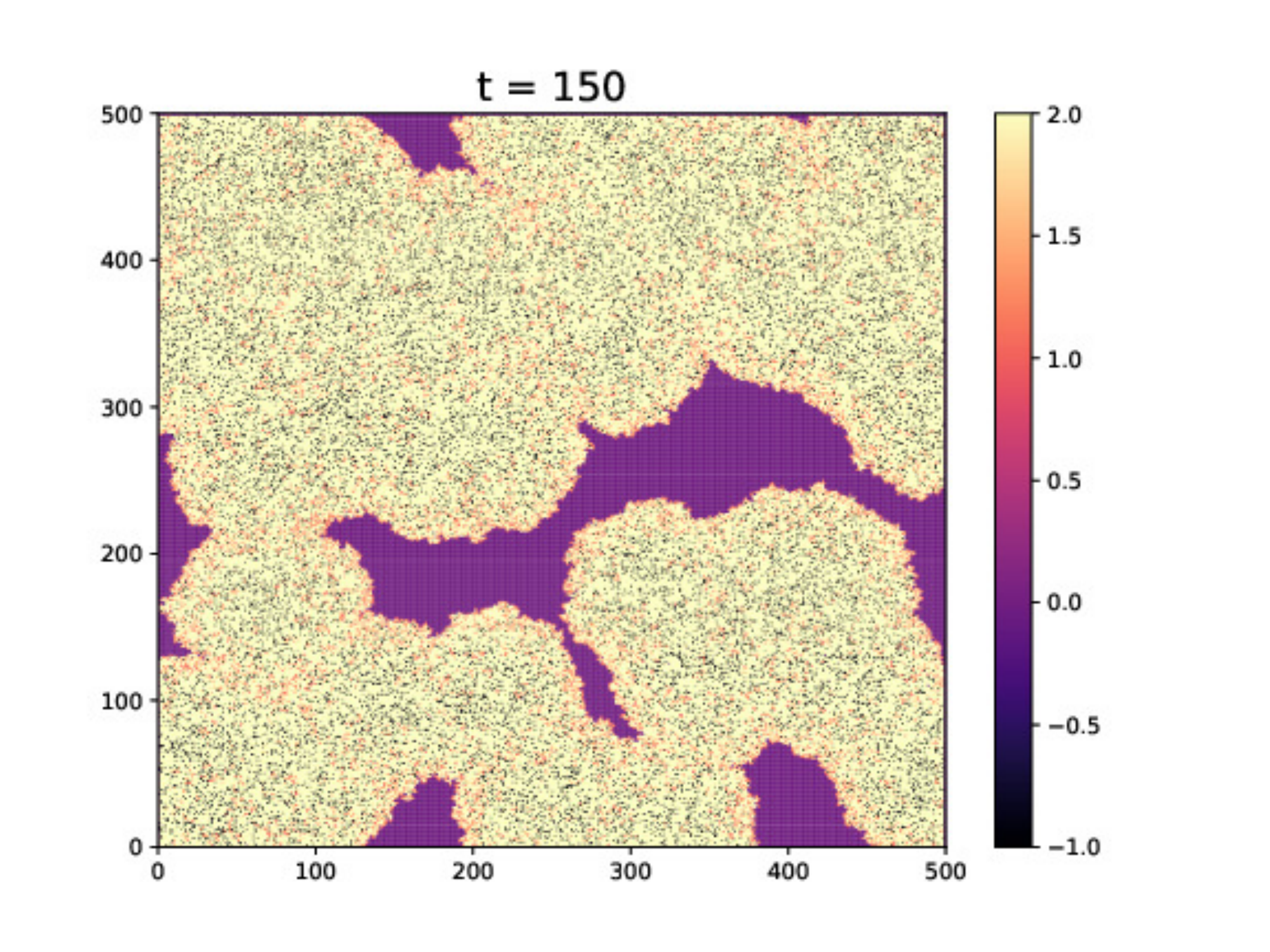}
 (d)\includegraphics[width=0.46\columnwidth]{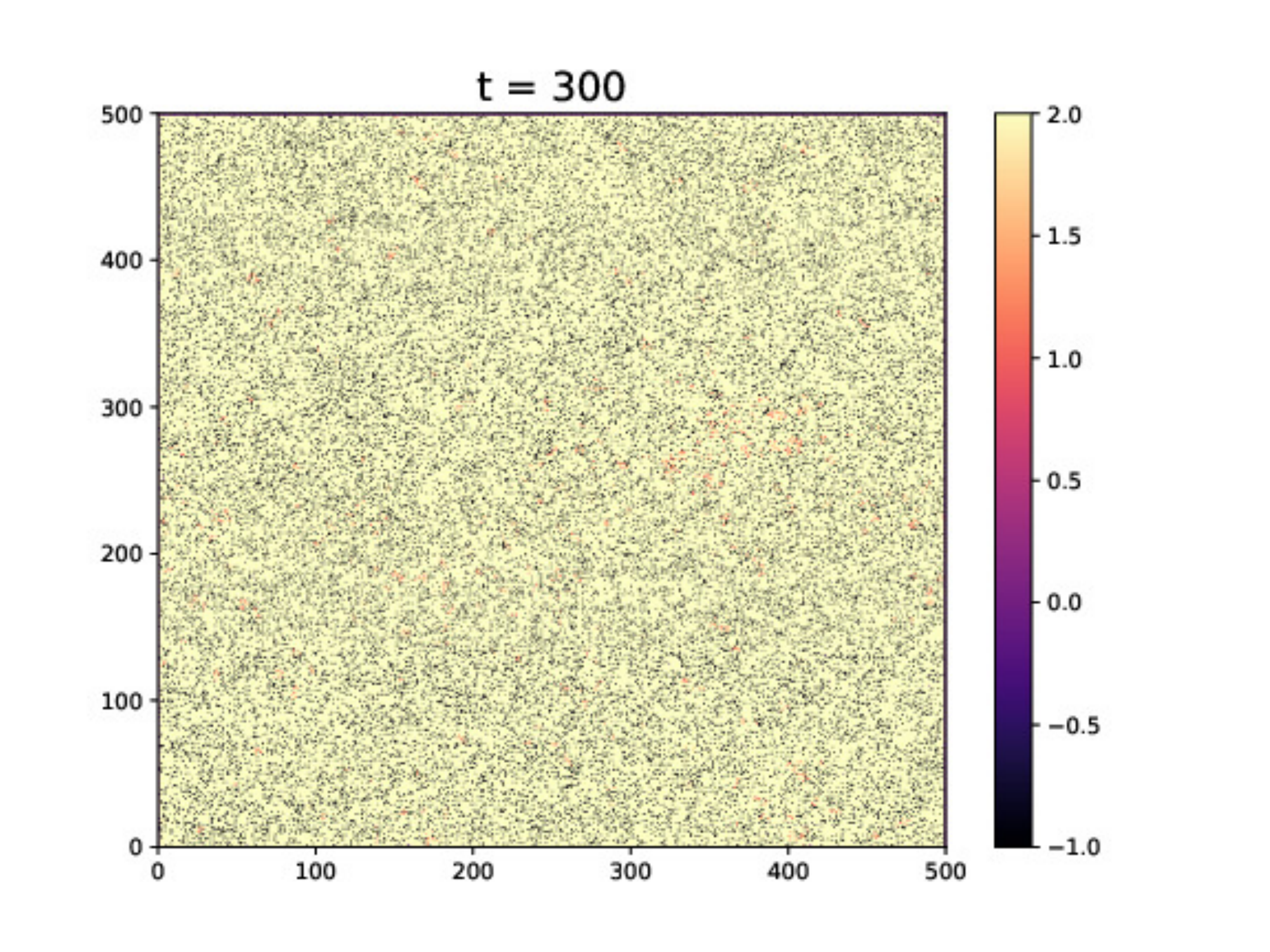}
 \caption{ (Color online) Chosen Parameters: Total population = 250000; total number of initially infected population = 50. The number -1 (black) represents the dead population; the number 0 (violet) represents the susceptible population; the range 1-2 (shades of red) represents the infected population; the number 2 (cream) represents the recovered population. Four frames of the Lattice Model Simulation in absence of vaccination are shown at    (a) $t=25$ (b) $t=100$ (c) $t=150$ (d) $t=300$. }
 \label{fig:500novac}
\end{figure*}

 FIG:~\ref{fig:100novac}  shows the 100x100 lattice at four different generations derived from Simulation 1: \textcolor{blue}{\url{https://youtu.be/raSCdZfT_5o}}. Similarly, FIG:~\ref{fig:500novac}  shows the 500x500 lattice at four different generations derived from Simulation 2: \textcolor{blue}{\url{https://youtu.be/zvjVY4Onhvw}}.

\subsection{In Presence of Vaccination}

\begin{figure}[h]
 \centering
 \includegraphics[width=0.88\columnwidth]{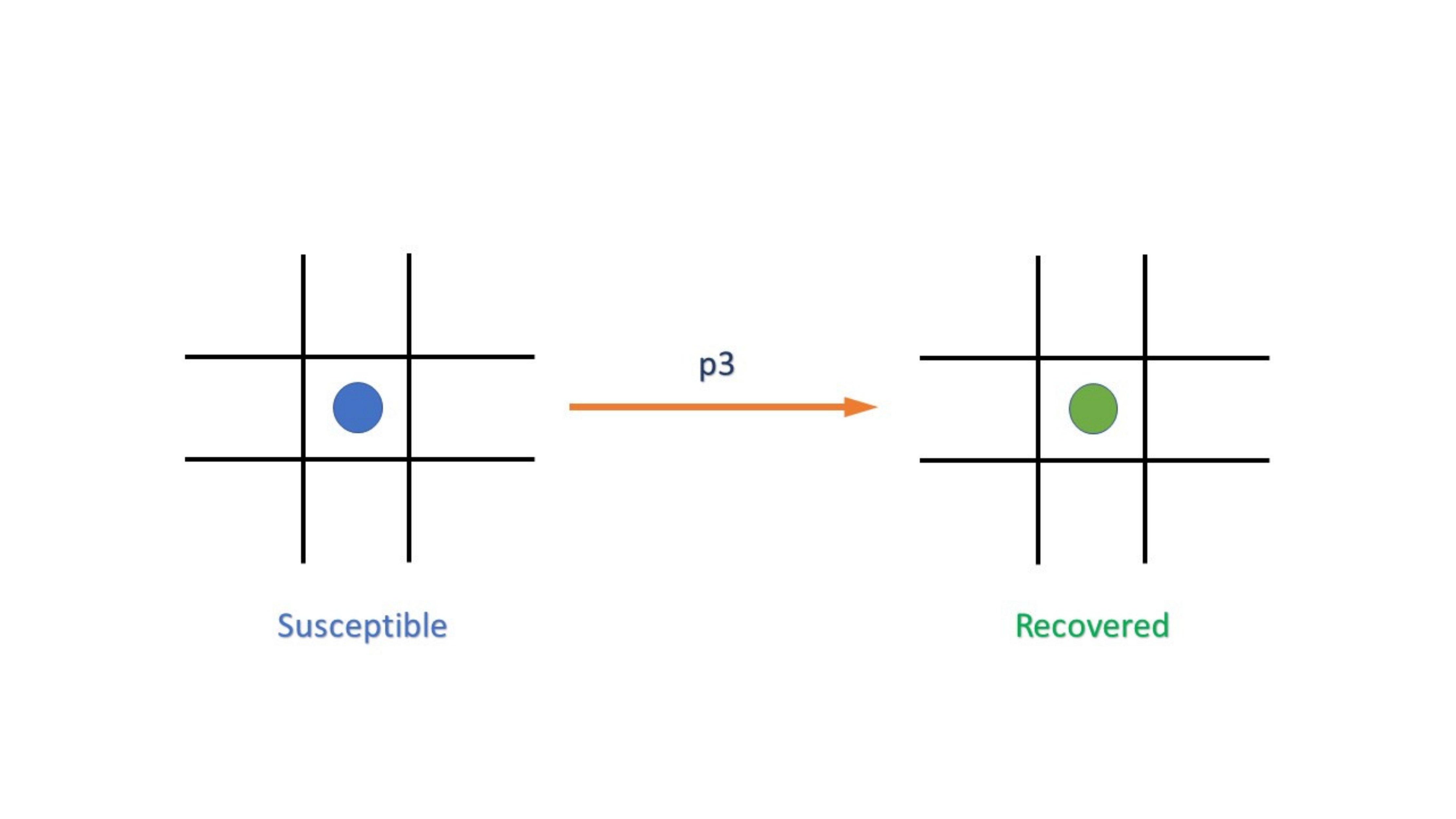}
 \caption{(Color online) In presence of vaccination, each susceptible cell has a probability p3 to get recovered directly without going through the infection phase when it can infect other cells.}
 \label{fig:vaccine}
\end{figure}

In this model, we follow the same three rules as in section 2.1. In addition to that, we introduce a new rule. Owing to the availability and effectiveness of the vaccine, we assign a probability p3 of getting vaccinated, i.e., getting converted to the recovered population, to each susceptible cell as shown in FIG:~\ref{fig:vaccine}. We execute this computationally as follows: \par \vskip 0.35cm

For any $t=t'$ and for all $i,j < N$, 
\par \vskip 0.35cm \hskip 0.5 cm
If $ A[i,j] = 0 $, set $r$ = random (generate a random number between 0 and 1). 
\par \vskip 0.25cm \hskip 0.5cm 
If $r< p3$, set $A[i,j]$ = 2, \hskip0.2cm  at $t=t'+1$. 

\vskip 0.35cm
Here also, we similarly run the simulation following the above rules for both  100x100 lattice and 500x500 lattice using the parameters as listed below in Table:~\ref{table2}. \par
\begin{figure*}[htpb]
 \centering
 (a)\includegraphics[width=0.46\columnwidth]{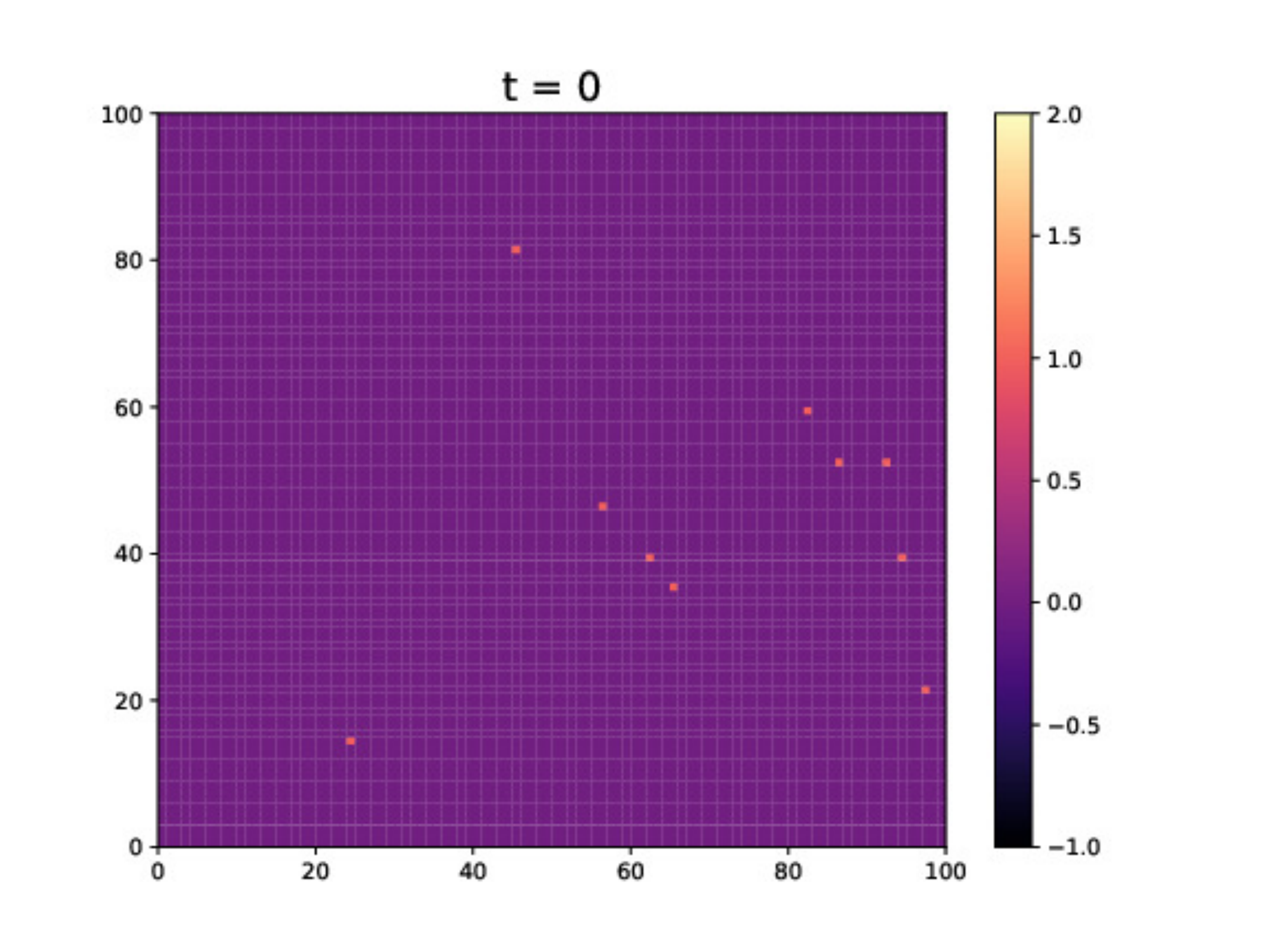}
 (b)\includegraphics[width=0.46\columnwidth]{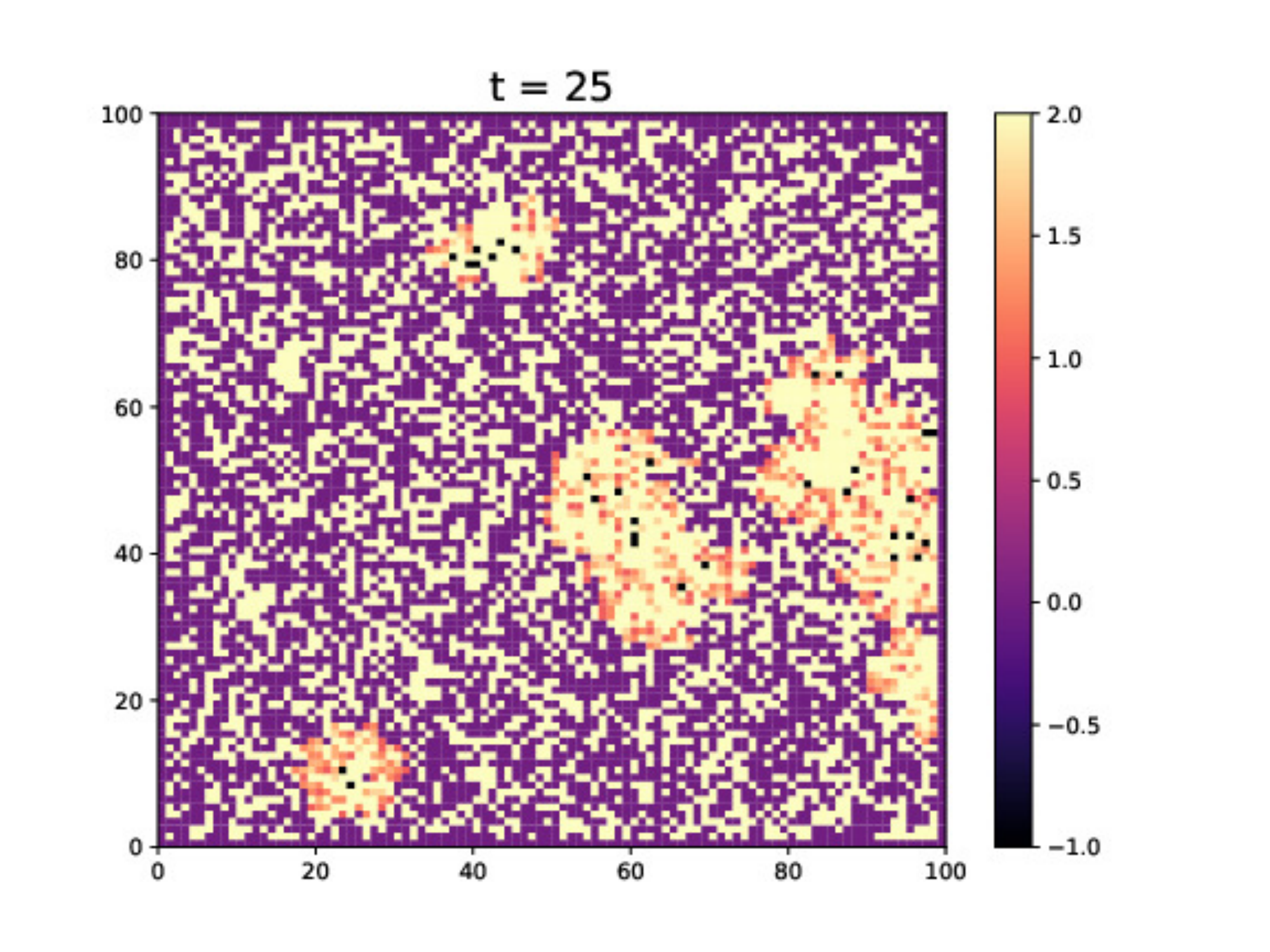}
 \\
 (c)\includegraphics[width=0.46\columnwidth]{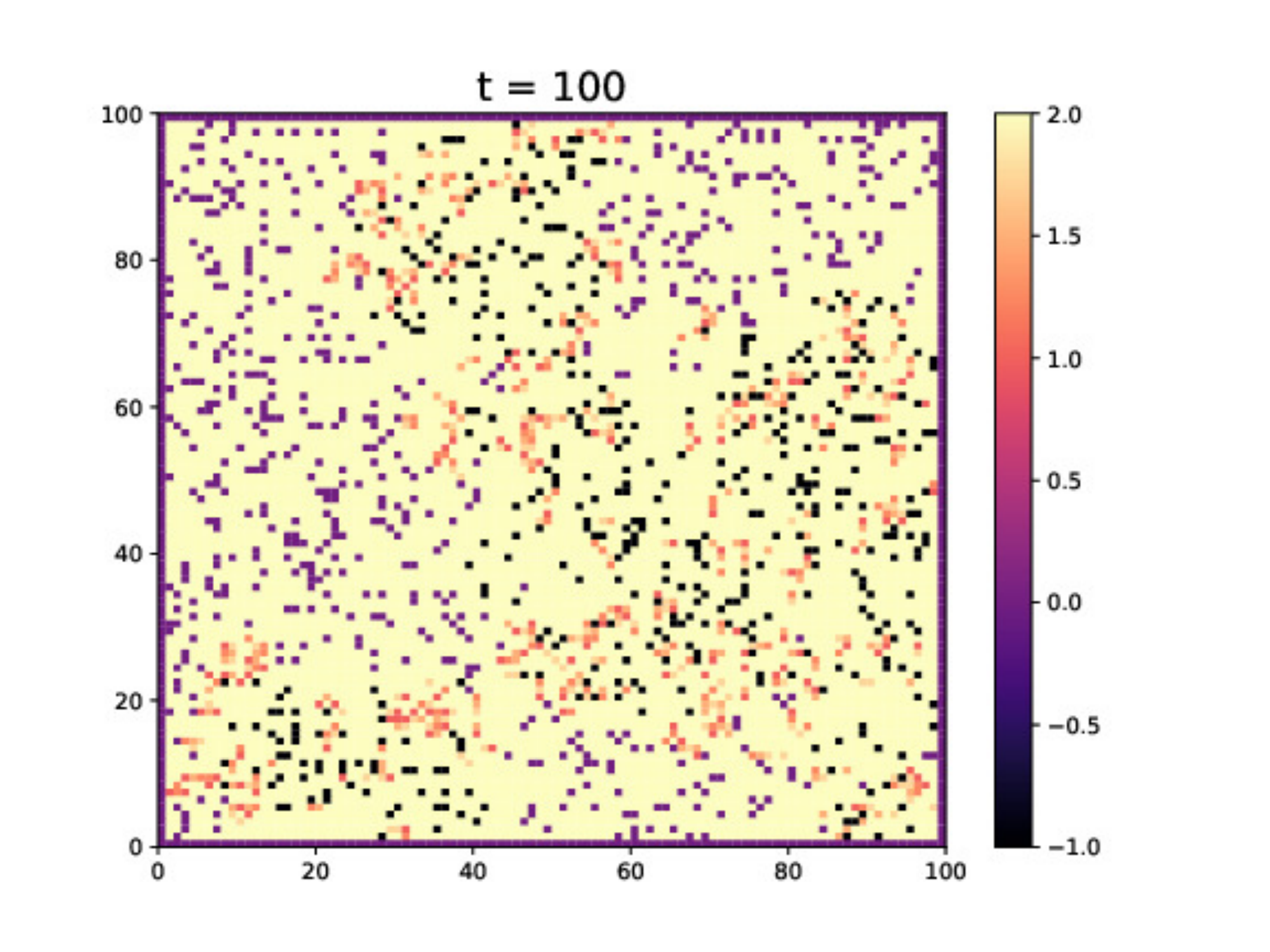}
 (d)\includegraphics[width=0.46\columnwidth]{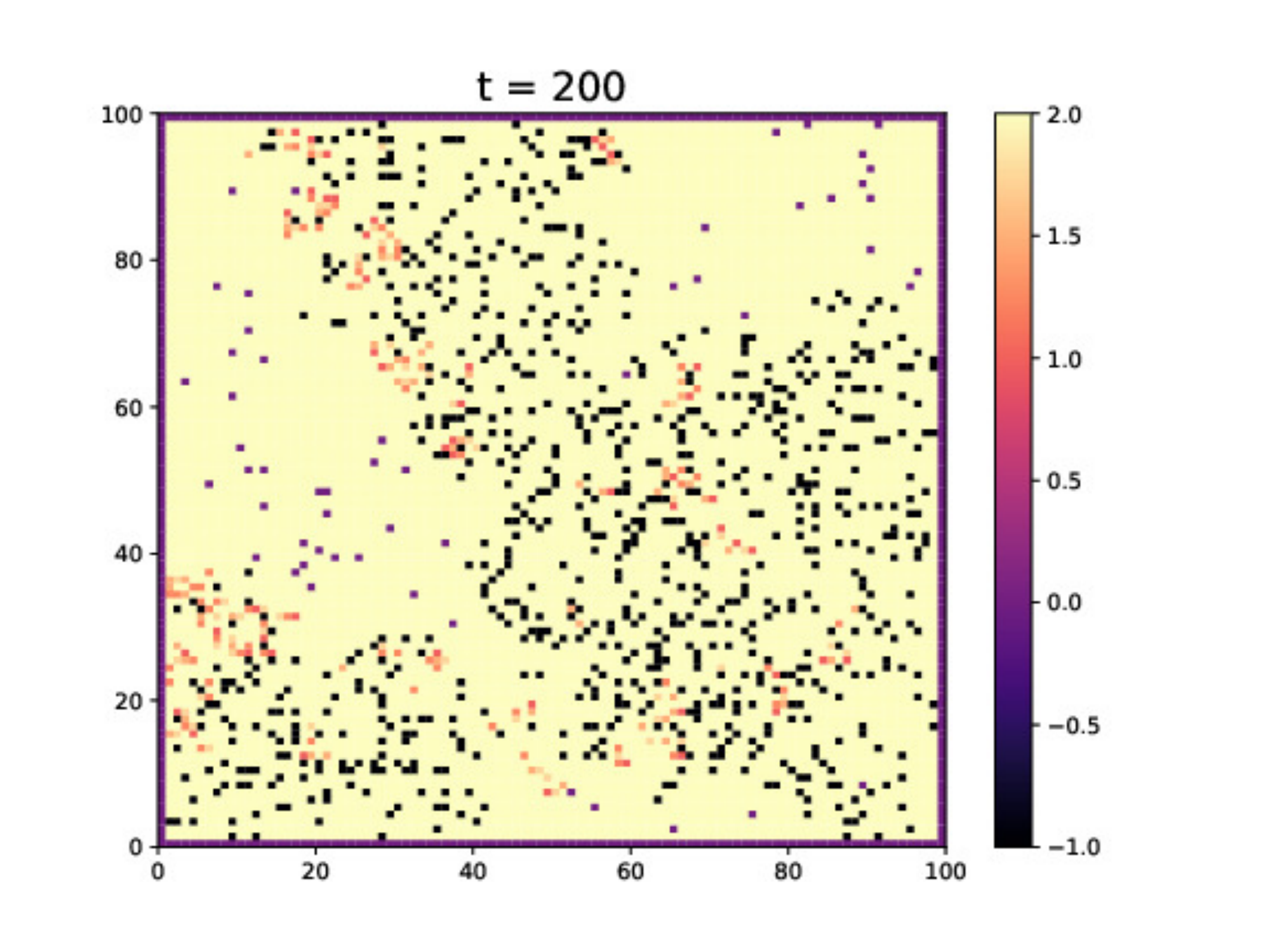}
 \caption{ (Color online) Chosen Parameters: Total population = 10000; total number of initially infected population = 10. The number -1 (black) represents the dead population; the number 0 (violet) represents the susceptible population; the range 1-2 (shades of red) represents the infected population; the number 2 (cream) represents the recovered population. Four frames of the Lattice Model Simulation in presence of vaccination are shown at    (a) $t=0$ (b) $t=25$ (c) $t=100$ (d) $t=200$. }
 \label{fig:100vac}
\end{figure*}

\begin{figure*}[htpb]
 \centering
 (a)\includegraphics[width=0.46\columnwidth]{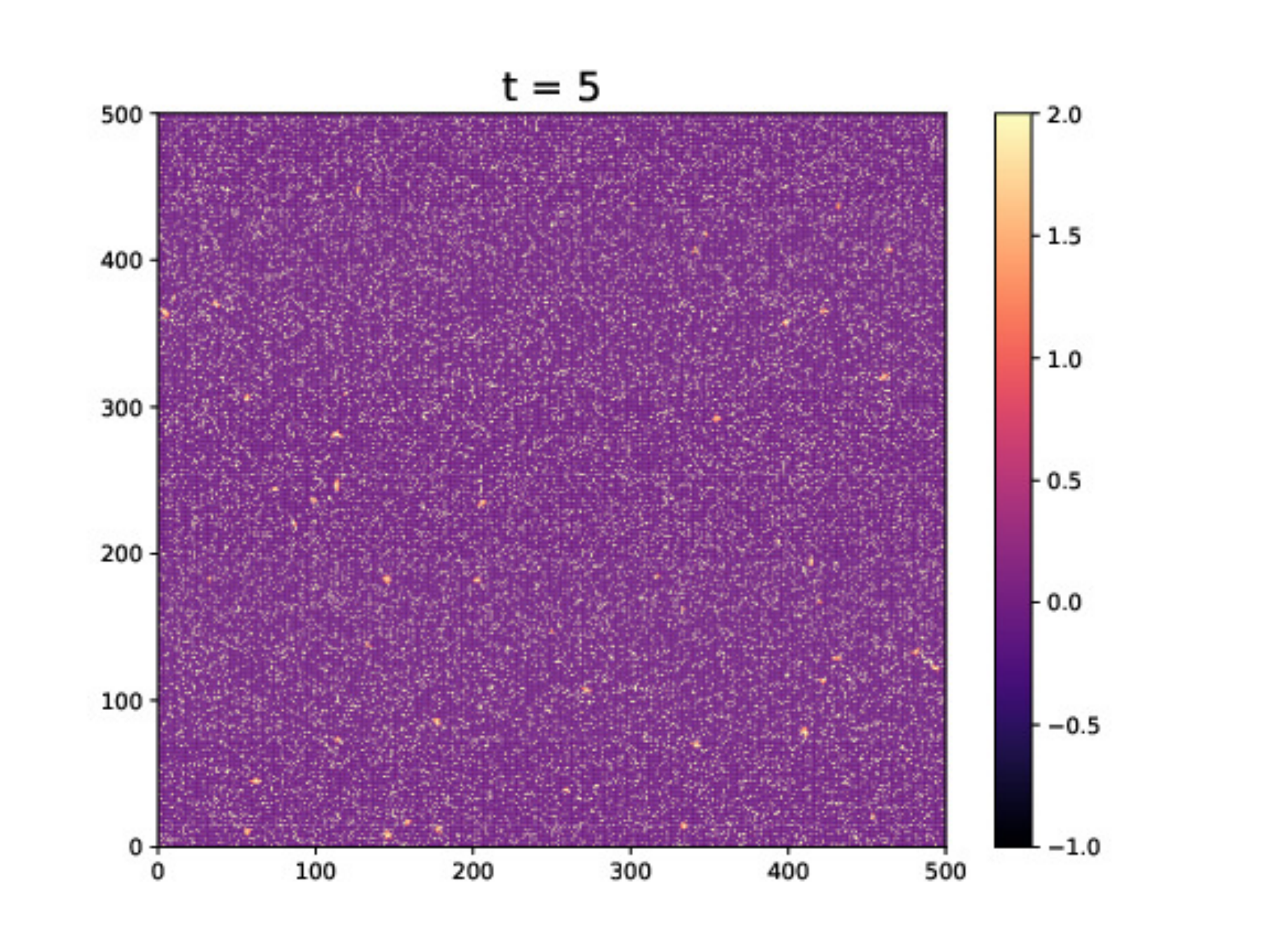}
 (b)\includegraphics[width=0.46\columnwidth]{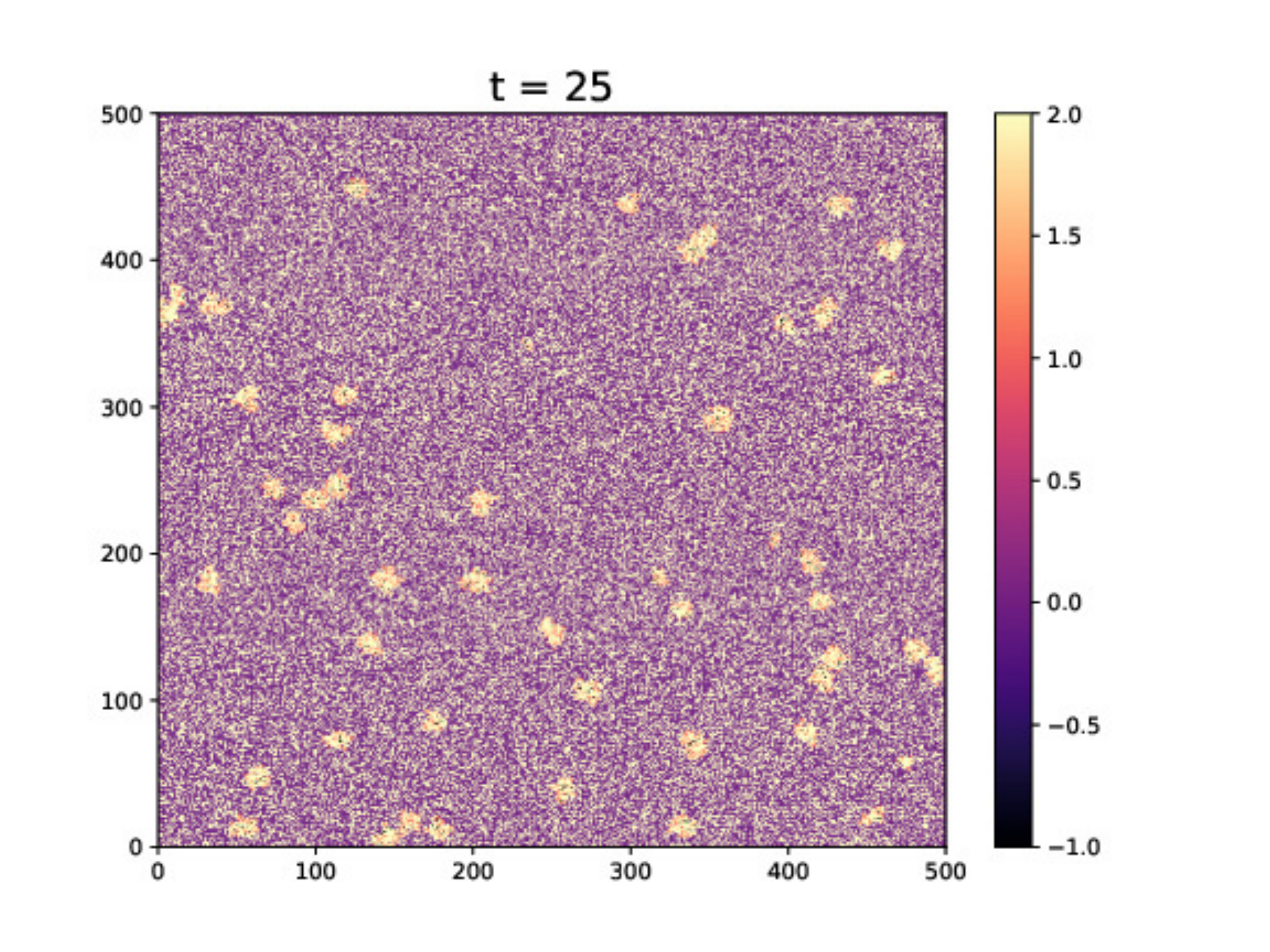}
 \\
 (c)\includegraphics[width=0.46\columnwidth]{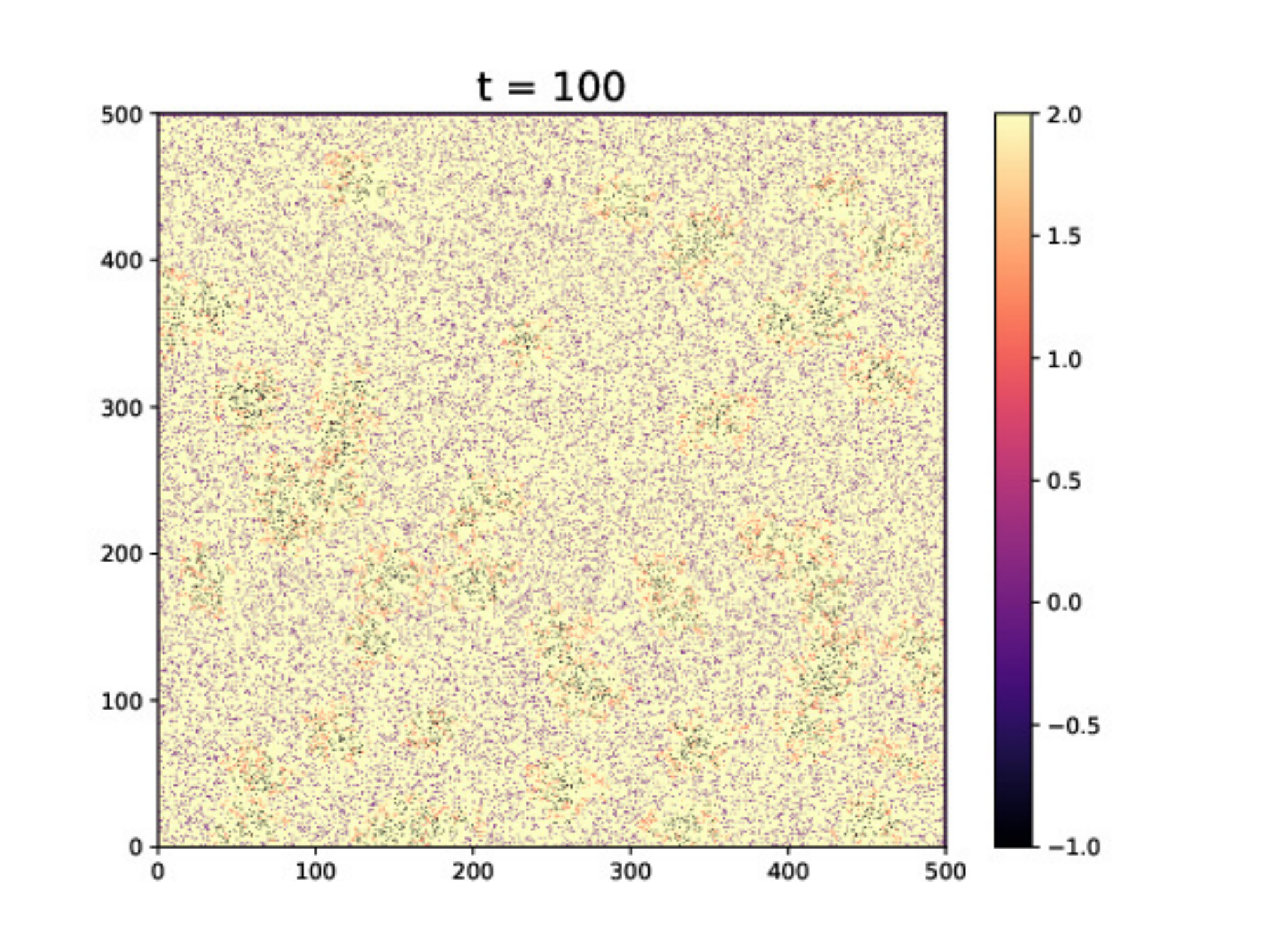}
 (d)\includegraphics[width=0.46\columnwidth]{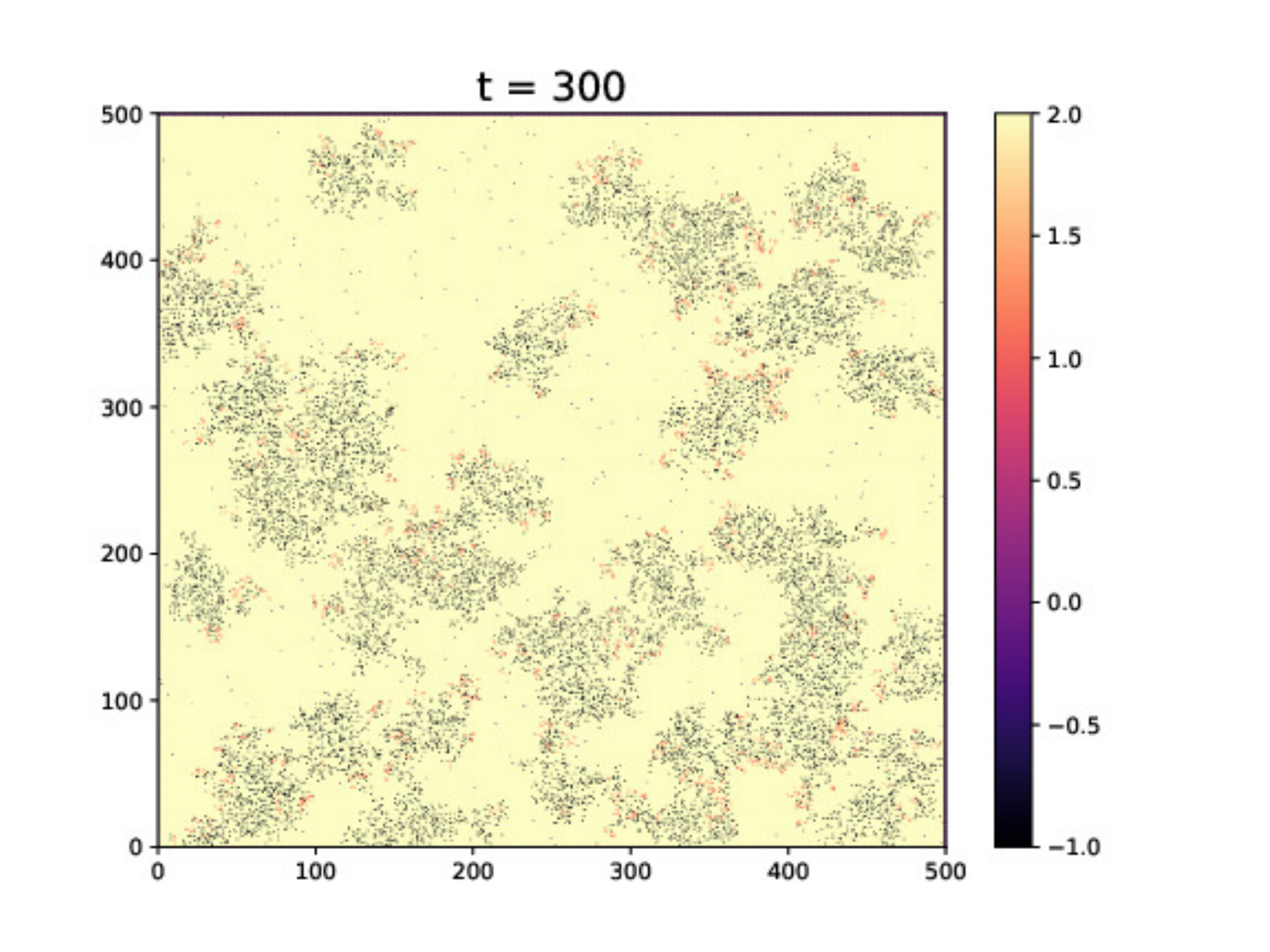}
 \caption{ (Color online) Chosen Parameters: Total population = 250000; total number of initially infected population = 50. The number -1 (black) represents the dead population; the number 0 (violet) represents the susceptible population; the range 1-2 (shades of red) represents the infected population; the number 2 (cream) represents the recovered population. Four frames of the Lattice Model Simulation in presence of vaccination are shown at    (a) $t=5$ (b) $t=25$ (c) $t=100$ (d) $t=300$. }
 \label{fig:500vac}
\end{figure*}

 FIG:~\ref{fig:100vac} shows the 100x100 lattice in presence of vaccination at four different generations derived from Simulation 3: \textcolor{blue}{\url{https://youtu.be/dJ2lsVhMq-Y}}. Similarly,FIG:~\ref{fig:500vac} shows the 500x500 lattice in presence of vaccination at four different generations derived from Simulation 4: \textcolor{blue}{\url{https://youtu.be/8O5-LYWNt9E}}.  

\section{Simulation Results and Discussion}
We analyse and discuss the results obtained from the simulations in subsection 2.1, i.e, in the absence of vaccination and subsection 2.2, i.e, in the presence of vaccination separately in the following two subsections.

\subsection{In Absence of Vaccination}

\begin{figure*}[htpb]
 \centering
 (a)\includegraphics[width=0.75\columnwidth]{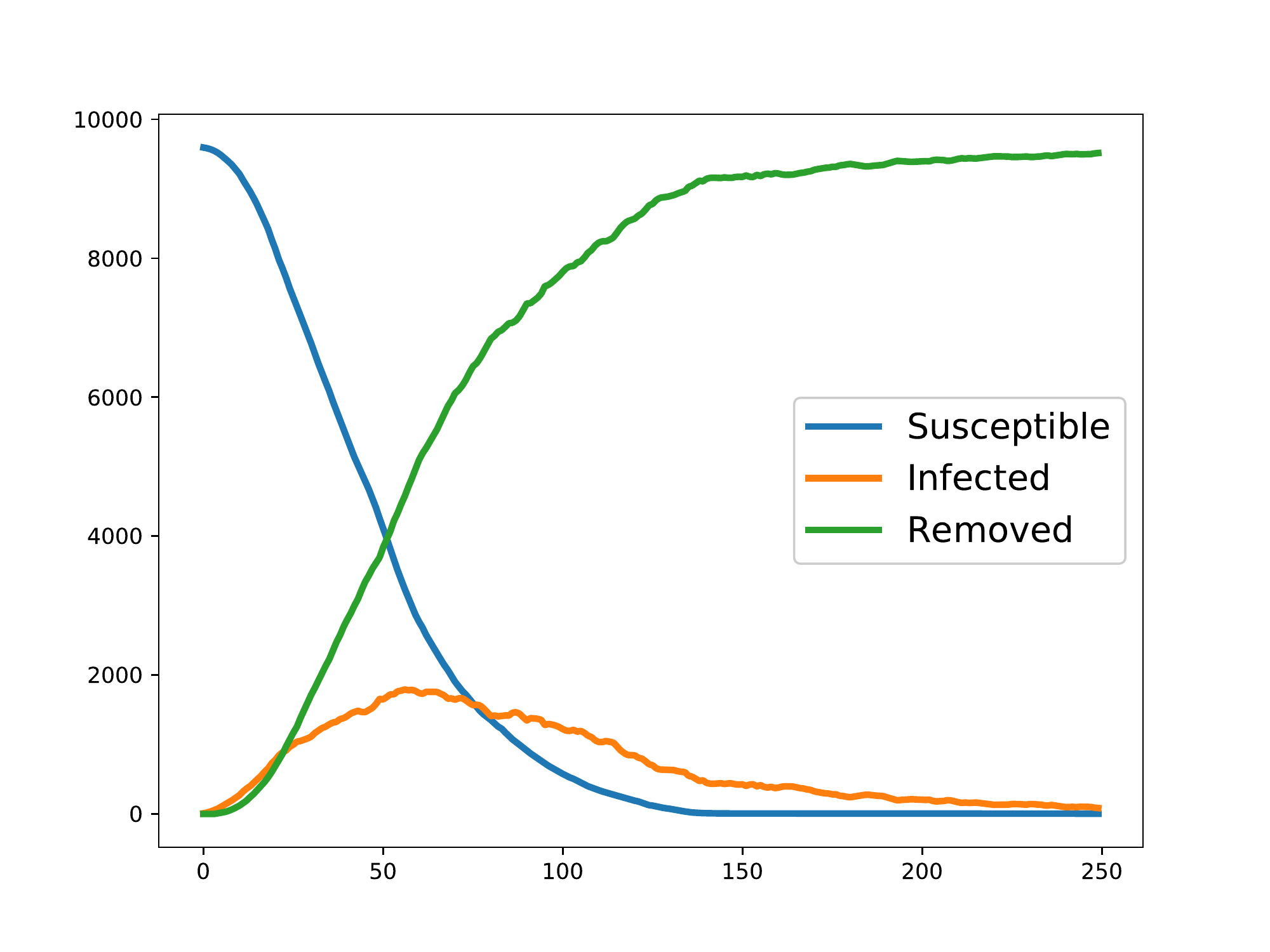}
 (b)\includegraphics[width=0.75\columnwidth]{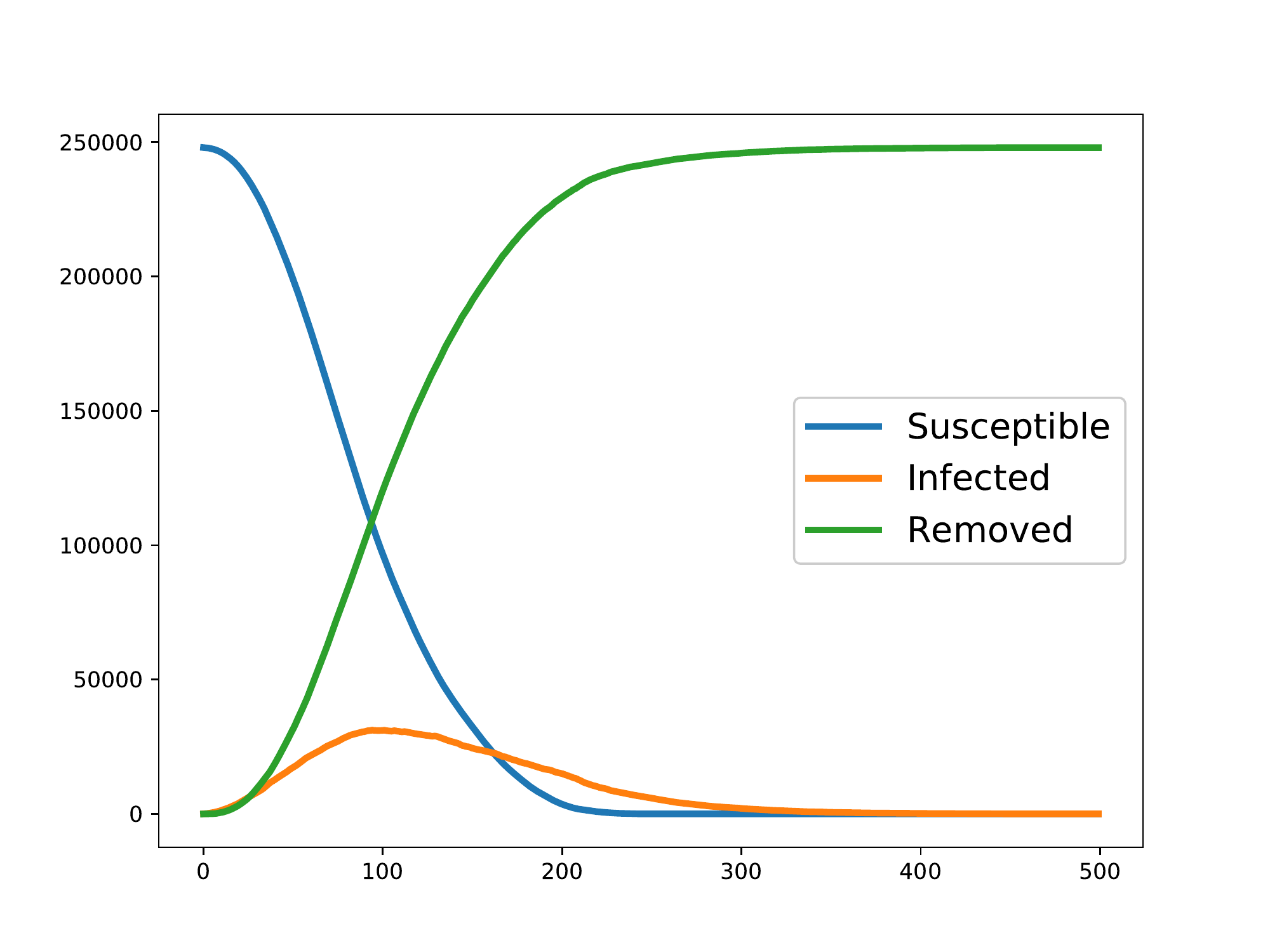}
 \\
 \caption{ (Color online) Plot of the counts of susceptible, infected and removed population at each time step in absence of vaccination for both (a)100x100 Lattice and (b)500x500 Lattice.} 
 \label{fig:novac}
\end{figure*}

While running simulations 1 and 2 as shown in FIG:~\ref{fig:100novac} and FIG:~\ref{fig:500novac}, we count the number of points having a value equal to 0 (susceptible), the number of points having a value in the range 1-2 (infected) and the total number of points having a value equal to 2 (recovered) and -1 (dead) at each time-step and plot them in both cases, as shown in FIG:~\ref{fig:novac}. We see that the susceptible population falls and then saturates with time; the infected population rises first to reach a peak and then falls down to zero with time; the removed (recovered and dead) population rises and saturates with time. These trends of the susceptible, infected and removed population for a typical epidemic match with that obtained from the SIR Model of epidemic proposed by Kermack and McKendrick in 1927 \cite{kermack}. From FIG:~\ref{fig:novac}, we also see that the fluctuations decrease as the system size increases. 

It may be noted here that if one considers the Moore neighbourhood of 2D cellular automata,
the positions of the peaks of infected populations would have shifted towards the left with a corresponding increase in height of the peaks of FIG:~\ref{fig:novac}.

\subsubsection{Velocity of Epidemic Spread}

\begin{figure*}[htpb]
 \centering
 (a)\includegraphics[width=0.46\columnwidth]{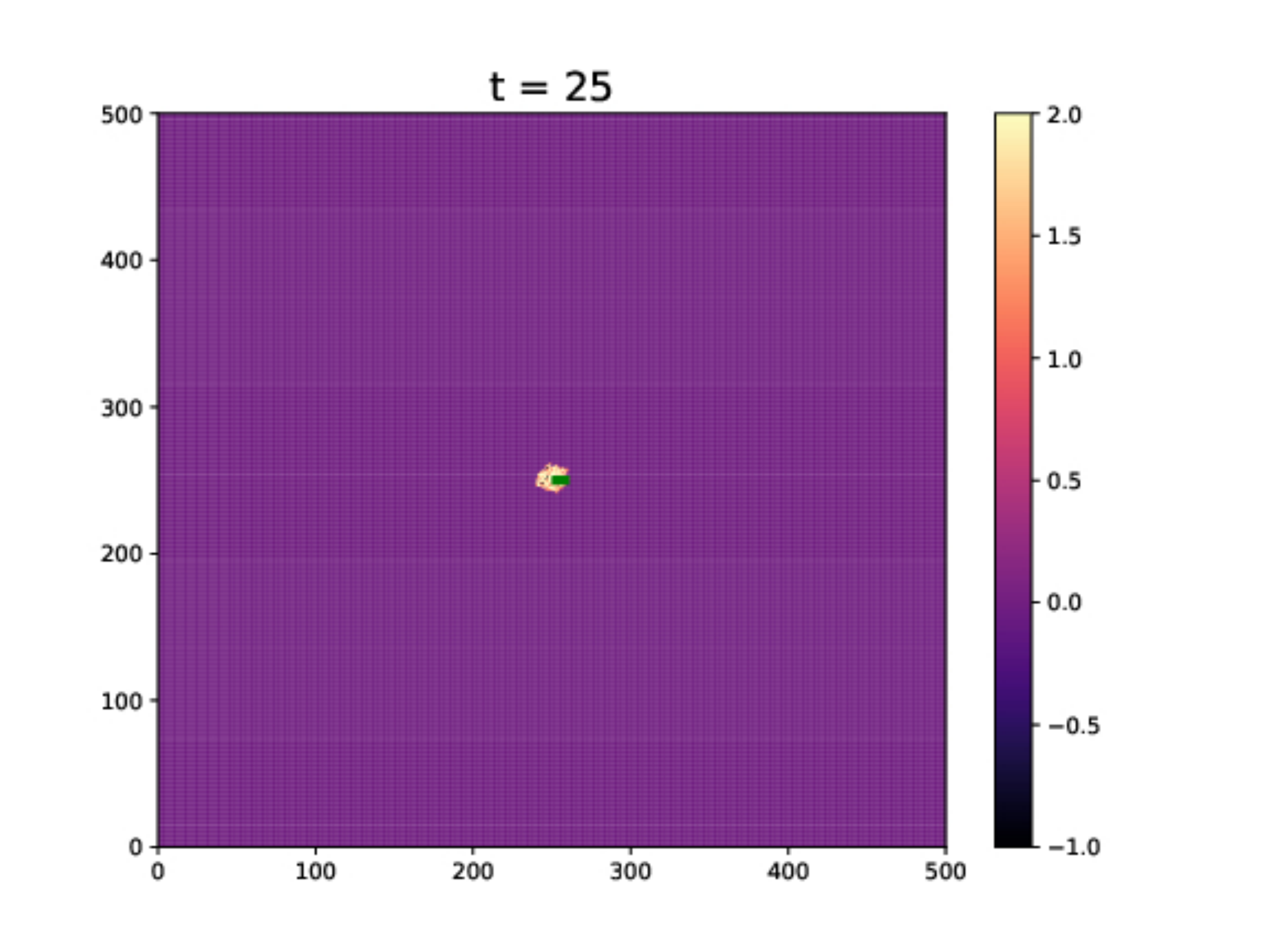}
 (b)\includegraphics[width=0.46\columnwidth]{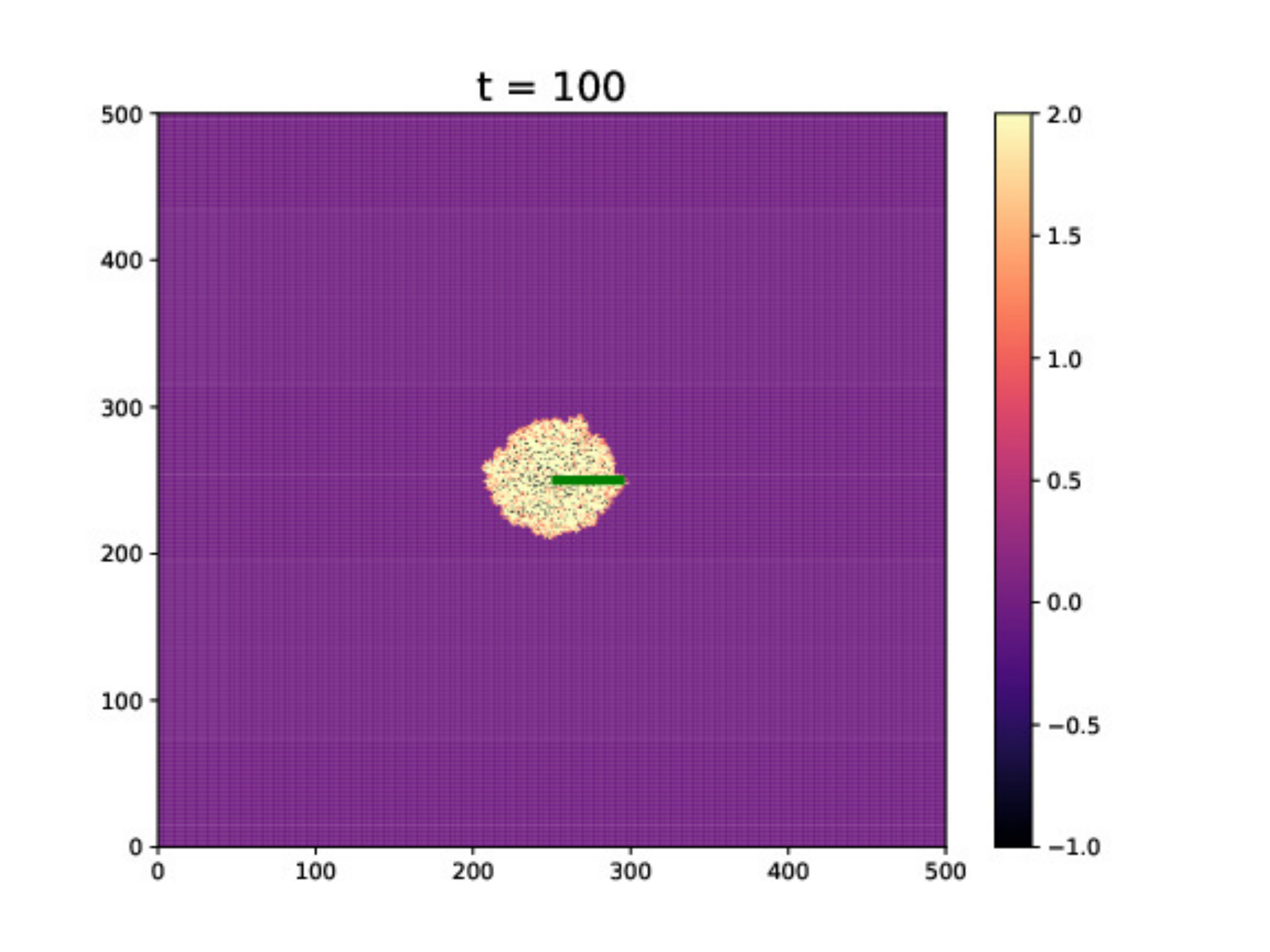}
 \\
 (c)\includegraphics[width=0.46\columnwidth]{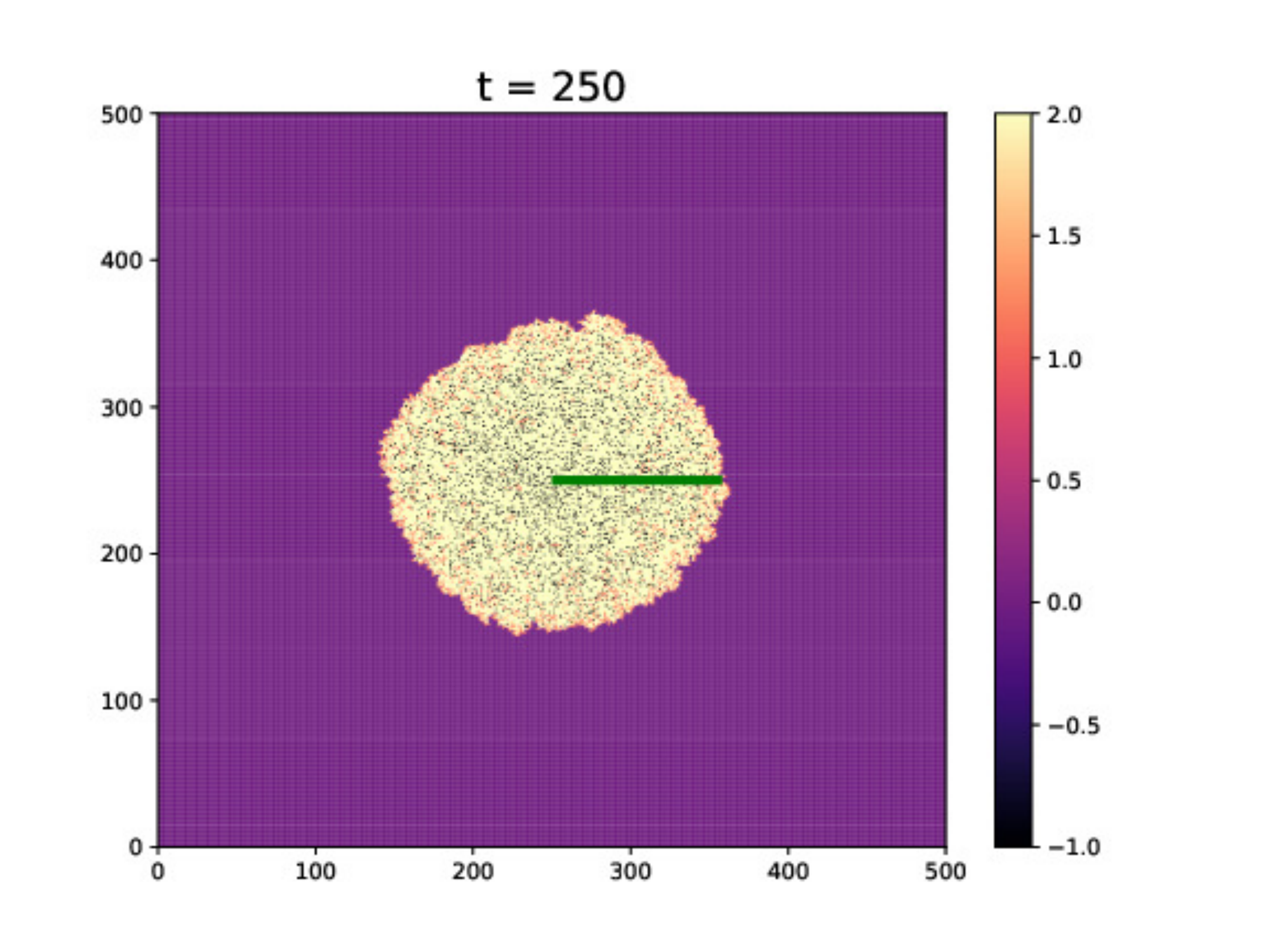}
 (d)\includegraphics[width=0.46\columnwidth]{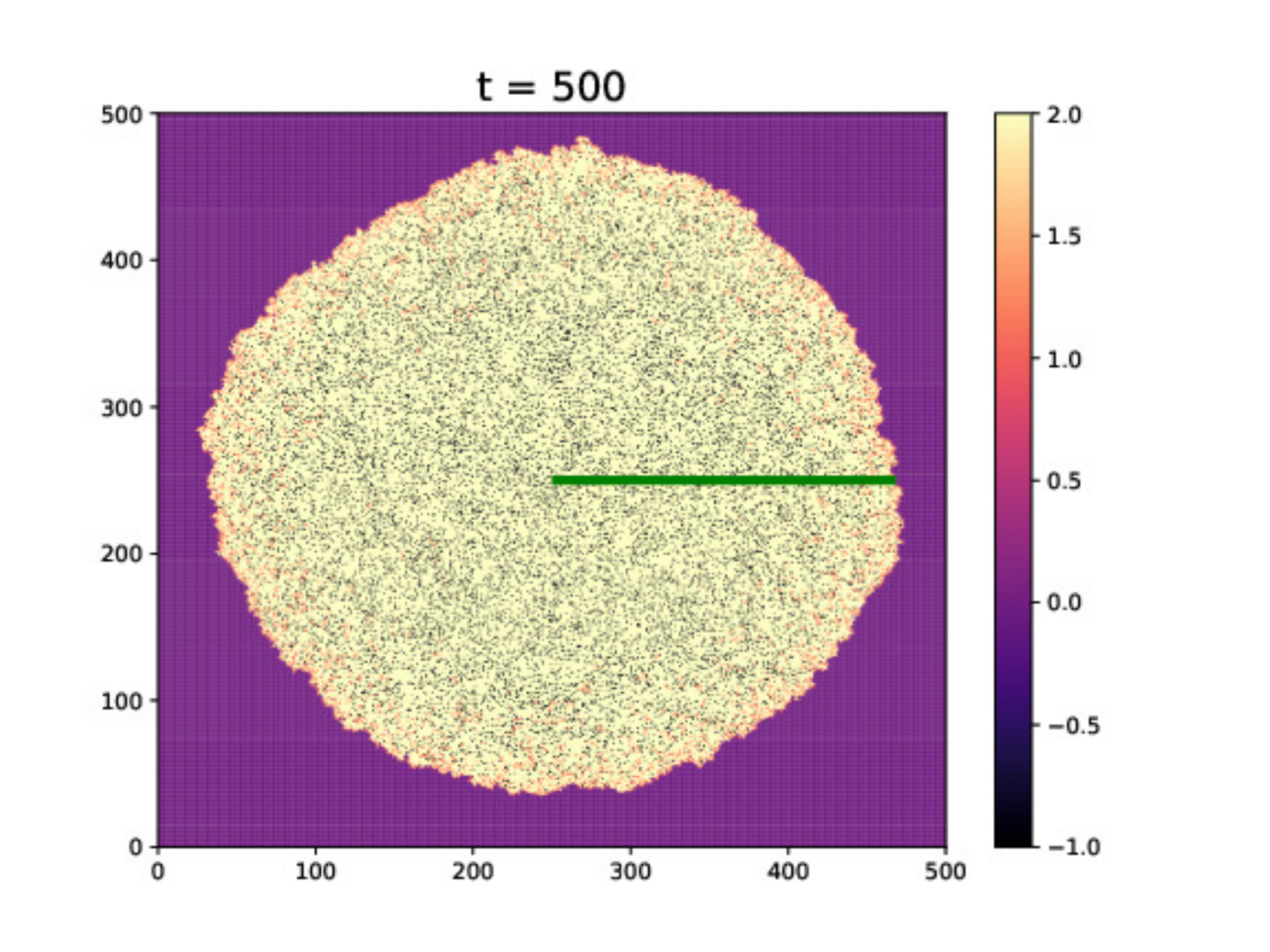}
 \caption{ (Color online) Chosen Parameters: Total population = 250000; total number of initially infected population = 1 (at the centre). The number -1 (black) represents the dead population; the number 0 (violet) represents the susceptible population; the range 1-2 (shades of red) represents the infected population; the number 2 (cream) represents the recovered population and the green line represents the radius of the infection bubble. Four frames of the Lattice Model Simulation are shown at (a) $t=25$ (b) $t=100$ (c) $t=250$ (d) $t=500$. }
 \label{fig:circle}
\end{figure*}

\begin{figure}[h]
 \centering
 \includegraphics[width=0.88\columnwidth]{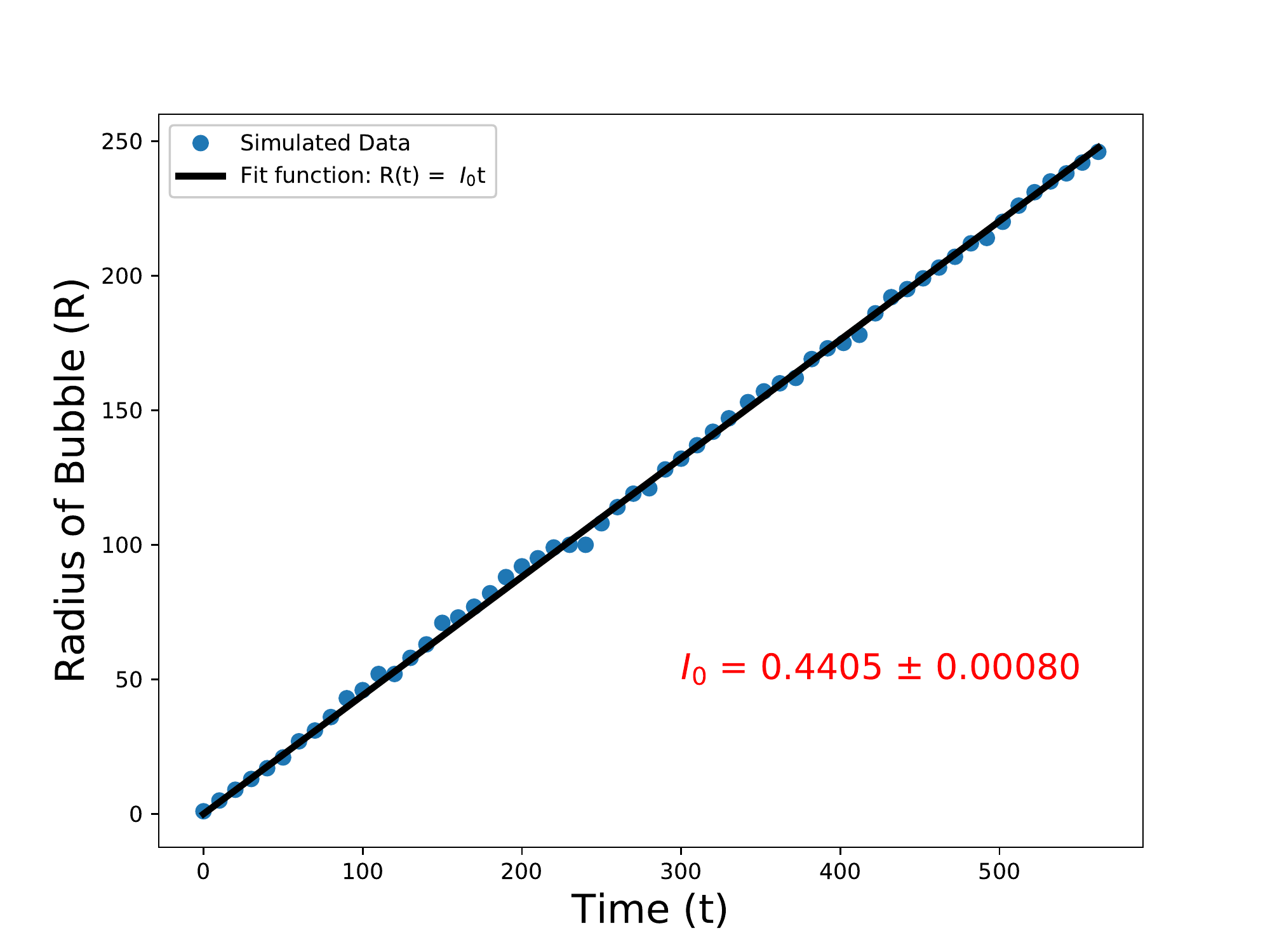}
 \caption{(Color online)Straight-line fit of the plot of radius of the infection bubble at each time step. }
 \label{fig:fit}
\end{figure}

We run a separate simulation using the rules from section 2.1 and the parameters from Table:~\ref{table1} considering only one infected cell exactly at the centre of a 500x500 lattice as given in Simulation 5: \textcolor{blue}{\url{https://youtu.be/Y1BUIA5BABo}}. FIG:~\ref{fig:circle} shows the 500x500 lattice at four different generations. We observe that the spread of infection is almost symmetric and hence circular. We plot the radius of the infection bubble $R$ at each time step(t) as shown in FIG:~\ref{fig:fit} and find,

\begin{equation}
    R \sim t
    \label{eqn:superdif}
\end{equation}

Now, if we obtained $R \sim \sqrt{t}$, we could have inferred that the nature of an epidemic spread is diffusive. But here the linear variation of $R$ with time tells us that the nature of an epidemic spread is essentially  superdiffusive. So, we find that the rate of spread of an epidemic is a constant of time. For our chosen parameters and initial condition, the value of the slope of the fitted straight line is 0.4405 $\pm$ 0.0008, which may be used to estimate the velocity of spread of different epidemics with proper scaling.

\subsubsection{Study of Correlation}

\begin{figure}[h]
 \centering
 \includegraphics[width=0.88\columnwidth]{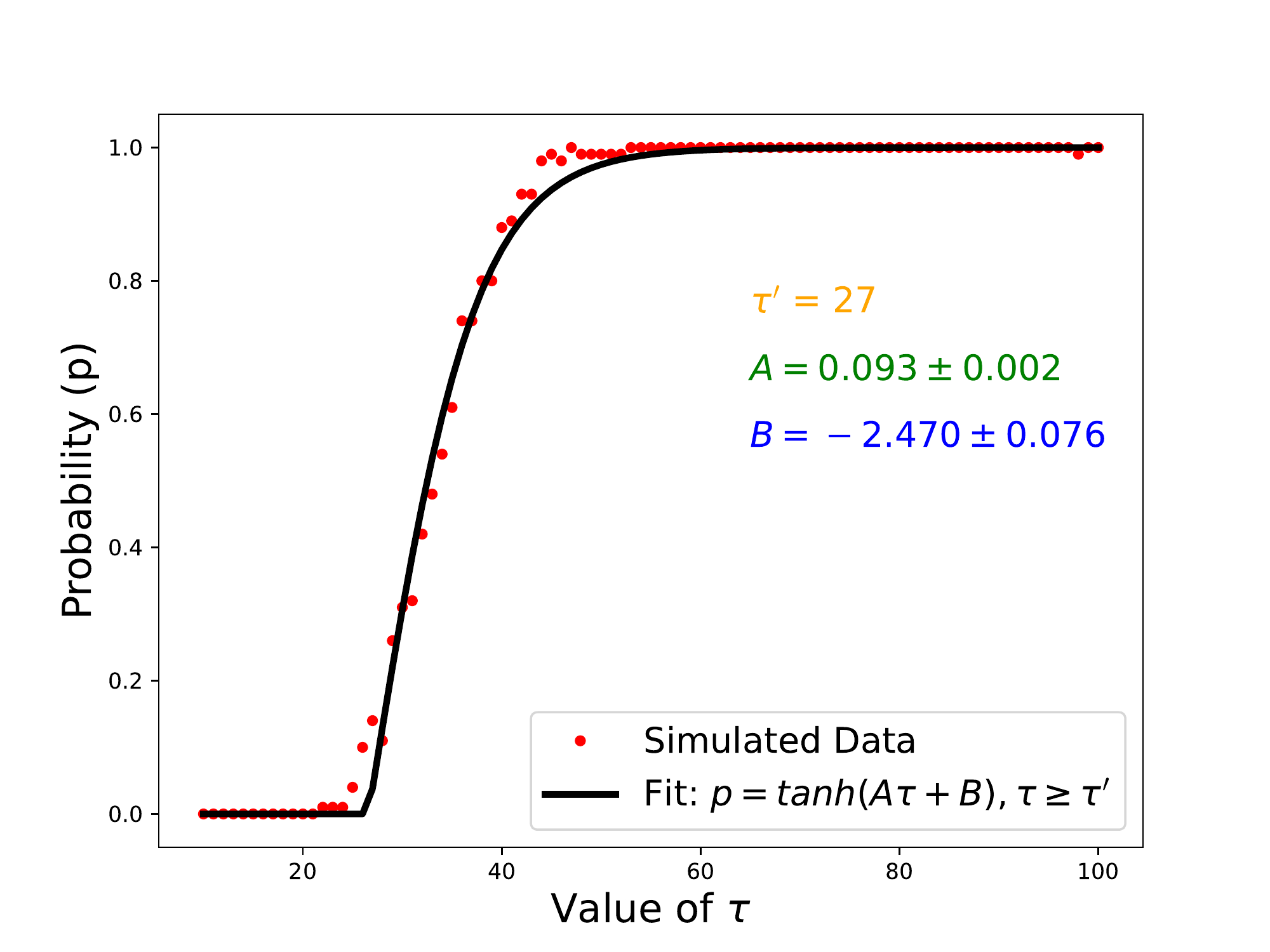}
 \caption{Plot to understand correlation of the central site with a particular site.}
 \label{fig:fitcor}
\end{figure}

We further study the correlation of two independent cells with respect to some initial conditions.
Here, one cell is considered as central between the two independent cells.
 We infect the central site in a background of a number of randomly oriented infected sites as the initial condition and then estimate the probability of a particular site in the lattice to be infected after $\tau$ time-steps or generations, by increasing the value of $\tau$, to understand the correlation between two infected sites. We execute this on a 100x100 lattice by infecting the central site in a background of 10 random infected sites. Now we start with $\tau = 10$ and run the simulation a number of times to observe whether a particular site gets infected or not, from where we calculate the probability of infection ($p$). We repeat this process by increasing the value of $\tau$. We plot the result in FIG:~\ref{fig:fitcor}, where we fit the obtained data in the functional form:

\begin{align}
p &= 0, \hskip2.5cm \tau < \tau'    \nonumber \\
\nonumber  \label{eqn:cor} \\
  &= tanh(A \tau + B), \hskip0.2cm \tau \geq \tau' \nonumber \\ 
\end{align}

From the fitted data, we estimate the values of the parameters to be $A = 0.093 \pm 0.002$ and $B = -2.470 \pm 0.076$. It is to be noted that the value $\tau'$ depends on the parameters $A$ and $B$, which ensures a non-zero value of the probability $p$, since the value of $tanh(A \tau + B)$ may be negative.    

It may be noted here that the correlation strongly depends on the choice of initial conditions. For example, if one considers the number of initially infected sites to be more than 10, the mean free space between two infected sites would decrease, as a result of which, the growth of correlation would have been faster. Moreover, as the distance between two chosen cells increases, the probability of infection in such correlation would decrease, leading to slower growth of correlation. This may lead to an interesting and valuable investigation.

\subsection{In Presence of Vaccination}

\begin{figure*}[htpb]
 \centering
 (a)\includegraphics[width=0.75\columnwidth]{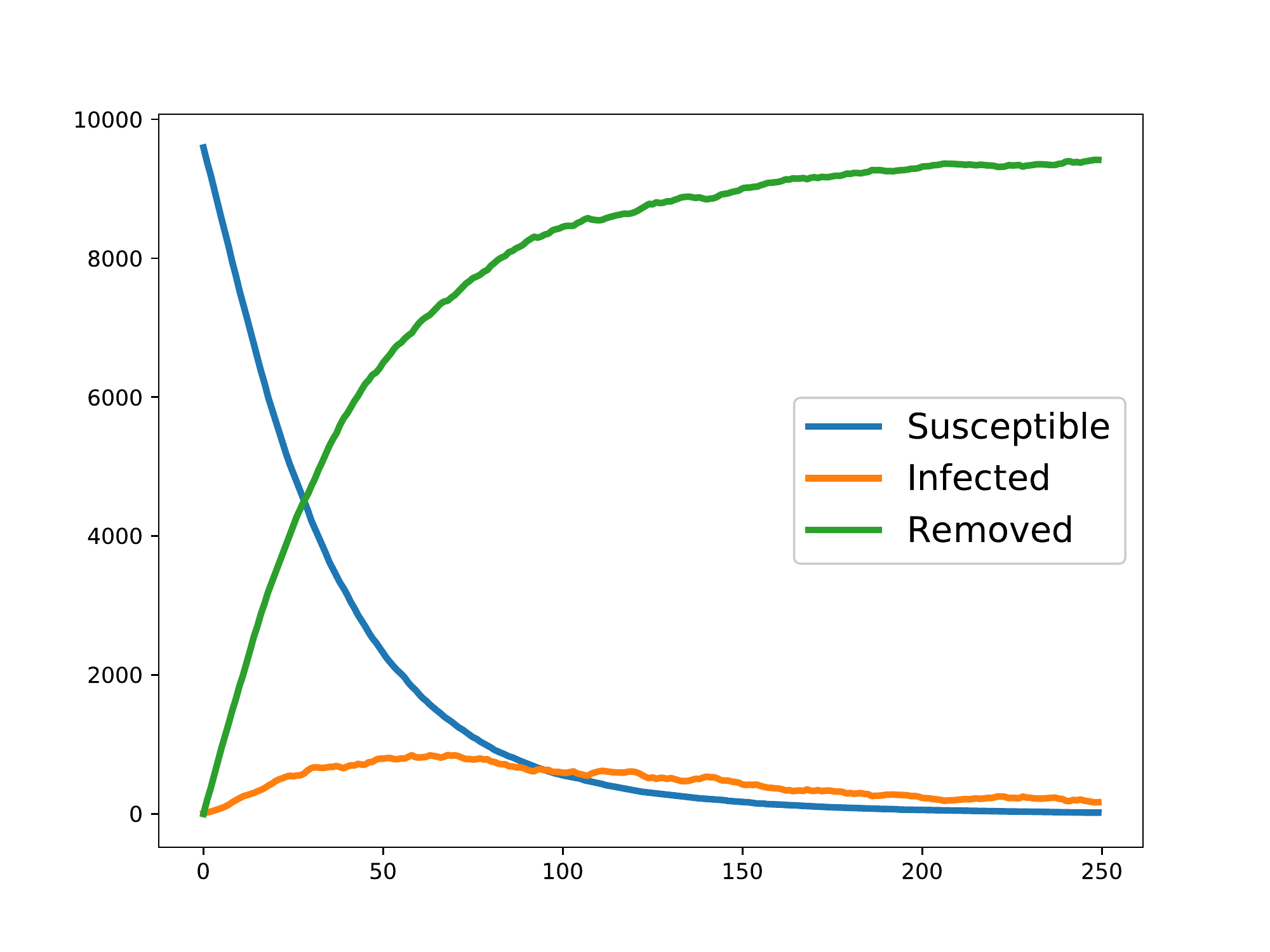}
 (b)\includegraphics[width=0.75\columnwidth]{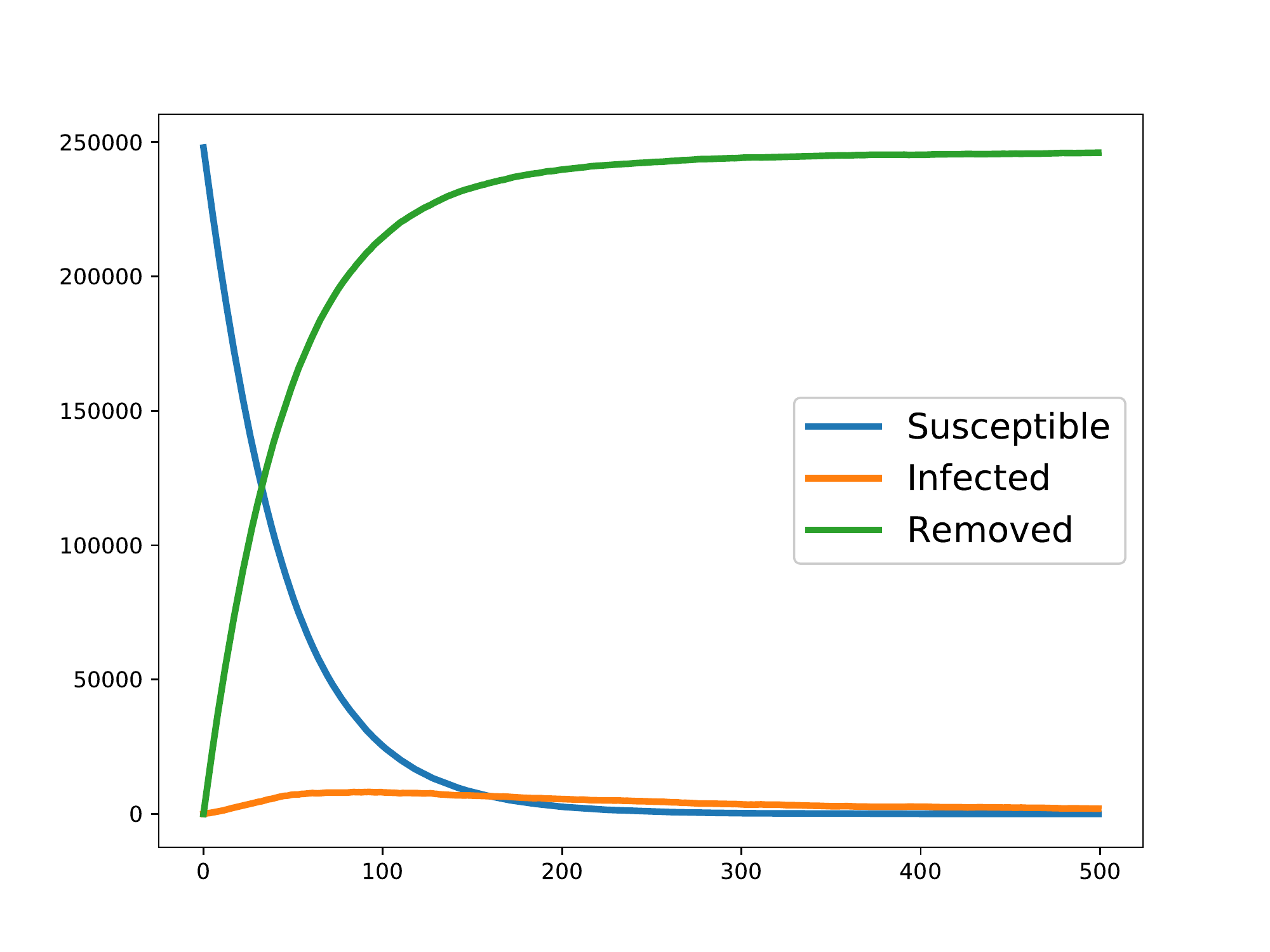}
 \\
 \caption{ (Color online) Plot of the counts of susceptible, infected and removed population at each time step in presence of vaccination for both (a)100x100 Lattice and (b)500x500 Lattice.} 
 \label{fig:vac}
\end{figure*}

\begin{figure}[h]
 \centering
 \includegraphics[width=0.88\columnwidth]{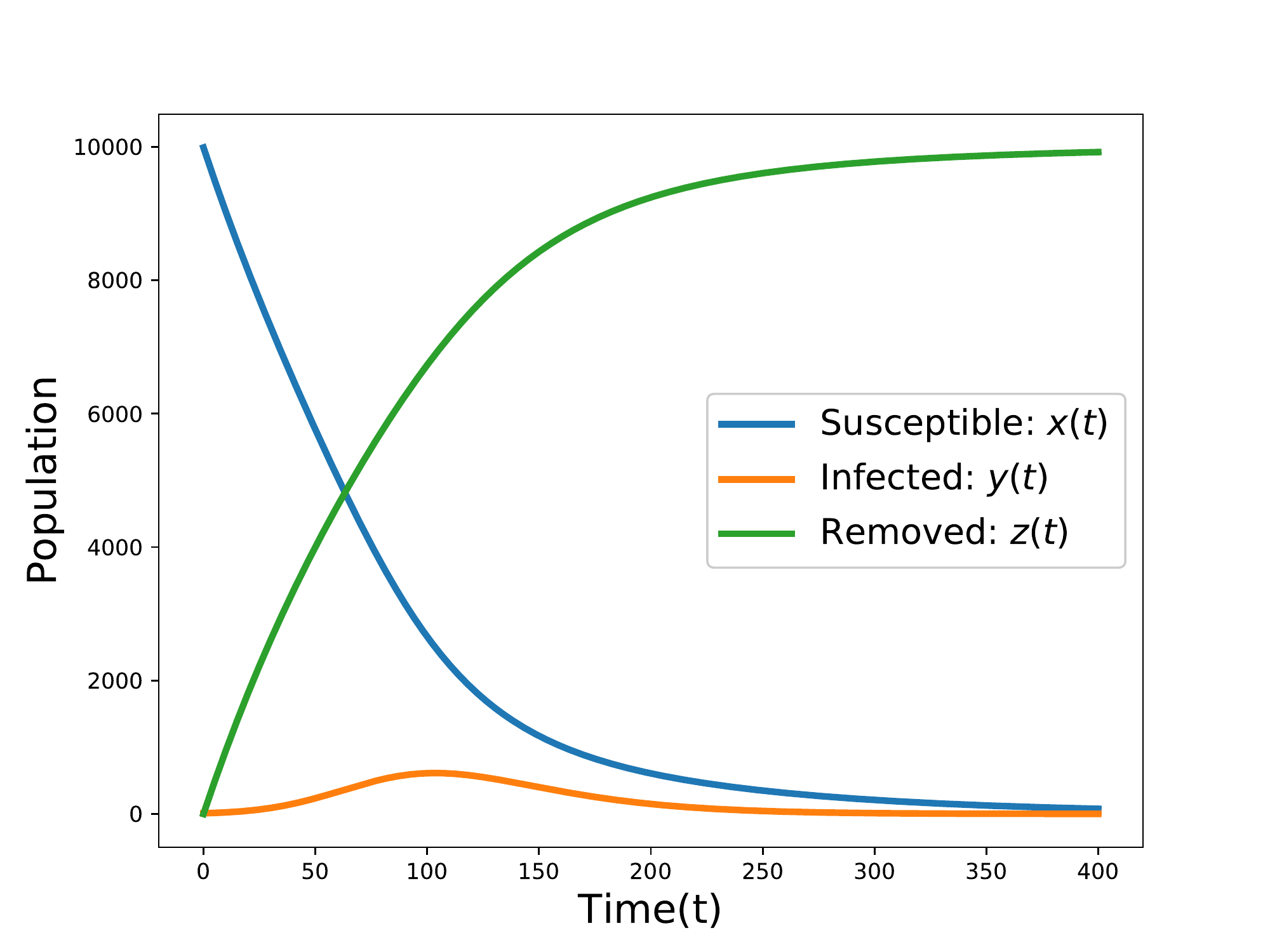}
 \caption{(Color online) Variation of susceptible, infected and removed population with time as predicted by Datta-Acharyya \cite{datta} }
 \label{fig:datta-acharyya}
\end{figure}

While running simulations 3 and 4 as shown in FIG:~\ref{fig:100vac} and FIG:~\ref{fig:500vac}, we similarly count and respectively plot the number of susceptible, infected and removed population at each time-step for both cases, as shown in FIG:~\ref{fig:vac}. We observe that the susceptible population falls almost exponentially with time; the infected population rises to reach a peak and then falls down to zero with time; the removed population saturates rapidly with time. These trends of the susceptible, infected and removed population for a typical epidemic match with that obtained from the Dynamical Model proposed by Datta-Acharyya \cite{datta}, which was developed as a modification to the original Kermack MCKendrick SIR Model in presence of medicated herd immunity, given by:

\begin{align}
\frac{dx}{dt} &= -kxy - cx     \nonumber \\
\nonumber \\
\frac{dy}{dt} &= kxy - ly  \label{eqn:vac} \\ 
\nonumber \\
\frac{dz}{dt} &= ly  + cx     \nonumber
\end{align}

where, $x$ = the number of susceptible people, $y$ = the number of infected people, $z$ = number of removed (recovered and dead) people, $k$, $l$ and $c$ are the three parameters which are the rate of infection, rate of removal and rate of vaccination respectively. We can solve Equation(1) numerically using the fourth-order Runge-Kutta Method and plot the variations of $x$, $y$ and $z$ with time for fixed values of $k$ and $l$, and simulate them for various values of $c$ to understand how an increase in the rate of vaccination can bring about significant change in the time variability of the three variables, as done by Datta-Acharyya in \textcolor{blue}{\url{https://youtu.be/JkmArmA-pC0}} \cite{datta}. We show one such plot in FIG:~\ref{fig:datta-acharyya}, from which we can see the similarity in trends of the variation of the susceptible, infected and removed population by comparing it to FIG:~\ref{fig:vac}. 

\vspace{0.3in}
\section{Conclusion} 

In this article, we have developed a lattice model of the spread of epidemic using cellular automata (CA). The CA algorithm has been framed in such a way that it can successfully capture the effects of getting infected even after recovery. The death
and survived-recovery have correctly been separated out in the original calculations (which was grossly mentioned in KM model as the removed population). The infection period has also been incorporated in the CA model. The basic target was to get the effects of the spatial evolution of the infection, which was not in Kermack-McKendrick mean-field kind of model. One can use this CA model in the extended system to get a more realistic picture of the spread of epidemic.

The basic results, regarding the time dependences of the number of Susceptible, Infected and Removed individuals, in the original KM model, are successfully reproduced in the CA model as a benchmark of the utility of this proposed CA model. The lattice model is used to show the growth of infection over the length space. The growth of
the dynamical shape of the infected cluster correctly reflects the symmetry of the
algorithm proposed. The multiple infected clusters grow and eventually coalesce to form a giant infected cluster. This giant infected cluster grows further to engulf the whole system. This picture of the growth of infection is observed in reality. 

{\it How does the infected cluster grow in time?}. This interesting question has also been addressed here and studied systematically (Fig-10). From the symmetry of the algorithm, the infected cluster is nearly circular in shape. The time-dependent radius of the infected cluster has been measured and plotted against time (Fig-11). Our results,
with good statistical analysis, fit in a form $R \sim t$.  The circular front of the infected cluster moves linearly in time. 

We further study the correlation of a particular site to be infected, given the central site is infected.  

Another important aspect of such SIR model is the inclusion of medicated immunity.
In the present CA model, the effects of medicated immunity have been incorporated
which agree quite well with the previously proposed model of the variant of the KM model.

Let us briefly mention the disadvantage of the model proposed in this study.
This CA model is described here on a square lattice in two dimensions. Consequently, the number of nearest neighbours are four. This restricts the saturation of the neighbourhood. This is a simplification of the model. If any model can be developed where the number of neighbours can be larger (e.g., in a network), one can employ the chances of infection spreading to the neighbours probabilistically and more efficiently, which would be more realistic as one person can infect more number of individuals due to various social contacts.

\vskip 1cm
\textbf{Acknowledgement}
AD thanks Presidency University for giving the opportunity to work in an academically engaging scientific environment. MA acknowledges FRPDF grant of Presidency University for financial support. We would like to thank Arunava Chakrabarti for his useful comments and suggestions.

\newpage 

\newpage

\begin{table}
    \centering
    \begin{tabular}{ |p{3.5cm}|p{3.5cm}| }
     \hline
     \multicolumn{2}{|c|}{Lattice Model without Vaccination} \\
     \hline
      Parameter & Value \\
     \hline
     p0 & 0.25  \\
        p1 & 0.97 \\
        p2 & 0.10  \\
     \hline
    \end{tabular}
    \caption{Table to show the parameters used to execute Simulation 1 and Simulation 2}
    \label{table1}
\end{table}

\begin{table}
    \centering
    \begin{tabular}{ |p{3.5cm}|p{3.5cm}| }
     \hline
     \multicolumn{2}{|c|}{Lattice Model with Vaccination} \\
     \hline
      Parameter & Value \\
     \hline
     p0 & 0.25  \\
        p1 & 0.97 \\
        p2 & 0.10  \\
        p3 & 0.02  \\
     \hline
    \end{tabular}
    \caption{Table to show the parameters used to execute Simulation 3 and Simulation 4}
    \label{table2}
\end{table}

\newpage

\newpage

\newpage

\newpage

\newpage

\newpage

\newpage

\newpage

\newpage

\newpage

\newpage

\newpage 

\newpage 

\newpage 

\end{document}